\documentclass[authoryear,preprint,12pt]{elsarticle}
\usepackage{setspace}
\usepackage{latexsym}
\usepackage{geometry}
\usepackage{bm}
\usepackage{boxedminipage}
\usepackage{amsmath}
\usepackage{graphicx}
\usepackage{graphics}
\usepackage{amsthm}
\usepackage{newlfont}
\usepackage{color}
\usepackage[normalem]{ulem}
\usepackage{subfigure}
\usepackage{enumerate}
\usepackage{float}
\usepackage{dsfont}
\usepackage{lineno}
\usepackage{amsfonts}
\usepackage{amssymb}
\usepackage{natbib}

\newcommand{\field}[1]{\mathbb{#1}}
\newcommand{\Z}{\field{Z}}

\newcommand{\E}{\field{E}}
\newcommand{\Var}{\field{V}ar}
\newcommand{\Cov}{\field{C}ov}
\renewcommand{\P}{\field{P}}
\newcommand{\etoile}[1]{#1^{\star}}
\newcommand{\Comb}[2]{\left(\begin{array}{c}#1 \cr #2\end{array}\right) }
\makeatletter

\@addtoreset{equation}{section} \makeatother

\newcommand{\1}[1]{\mathds{1}_{\left \lbrace #1\right\rbrace}}

\newtheorem{prop}{Proposition}
\makeatletter

\@addtoreset{prop}{section} \makeatother

\graphicspath{{./Graph/}}
\journal{Computational Statistics and Data Analysis }

\begin{document}
\begin{frontmatter}
\title{\Large Random effects compound Poisson model to represent data with
extra zeros}
{\author[l1,l2]{Marie-Pierre \'Etienne \corref{cor1} }
\author[l1,l2]{\'Eric Parent}
\author[l3]{Hugues Benoit}
\author[l1]{Jacques Bernier}
}
\cortext[cor1]{
\ead[url]{www.agroparistech.fr/morse/etienne.html}
\ead{marie.etienne@agroparistech.fr}
D\'epartment MMIP - Team Morse. 16 rue Claude Bernard, 75231 Paris Cedex 5. FRANCE
}

\address[l1]{AgroParisTech, UMR 518, F-75000 Paris, France}
\address[l2]{INRA, UMR 518, F-75000 Paris, France}
\address[l3]{Fisheries and Oceans Canada, Moncton, New Brunswick,Canada}
\date{\today}
\begin{abstract}
This paper describes a compound Poisson-based random effects
structure for modeling zero-inflated data. Data with large
proportion of zeros are found in many fields of applied statistics,
for example in ecology when trying to model and predict species
counts (discrete data) or abundance distributions (continuous data).
Standard methods for modeling such data include mixture and two-part
conditional models. Conversely to these methods, the stochastic
models proposed here behave coherently with regards to a change of
scale, since they mimic the harvesting of a marked Poisson process
in the modeling steps. Random effects are used to account for
inhomogeneity. In this paper, model design and inference both rely
on conditional thinking to understand the links between various
layers of quantities~: parameters, latent variables including random
effects and  zero-inflated observations. The potential of these
parsimonious hierarchical models for zero-inflated data is
exemplified using two marine macroinvertebrate abundance datasets
from a large scale scientific bottom-trawl survey. The  EM algorithm
with a Monte Carlo step based on importance sampling is checked for
this model structure on a simulated dataset~: it proves to work well
for parameter estimation but parameter values matter when
re-assessing the actual coverage level of the confidence regions far
from the asymptotic conditions.
\end{abstract}
\end{frontmatter}

\linenumbers

\section{Introduction}
 Often data contain a greater number of zero
observations than would be predicted using standard, unimodal
statistical distributions. This currently happens in ecology (see
\cite{Martinetal2005}) when counting species (over-dispersion for
discrete data) or recording biomasses (atoms at zero for continuous
data). Such data are generally referred to as zero-inflated data and
require specialized treatments for statistical analysis
\citep{Heilbron+94}. Common statistical approaches to modeling
zero-inflated data make recourse either to mixture models, such as
the Dirac function for the occurrence of extra zeros in addition to
a standard probability distribution (see for instance
\cite{Ridout+98}), or to two-part conditional models (a
presence/absence Bernoulli component and some other distribution for
non zero observations given presence such as in
\cite{Stefansson+96}). These models are well-known
\citep{BarryWelsh2002} and offer the advantages of separate fits and
separate interpretations of each of their components. Parameters are
well understood and interpreted as the probability of presence, and
the average abundance of biomass if present.

However, a major flaw of those models is their non-additive behavior
with regards to variation in within-experiment sampling effort
\citep{Syrjala+2000}. Consider for instance the fishing effort
measured by the ground surface swept by a bottom-trawl during a
scientific survey of benthic marine fauna. If during experiment $i$,
observation $Y_{i}$ is made with some experimental effort
corresponding to the harvesting of some area $D_{i}$ and is assumed
to stem from a stochastic model with parameters $\theta(D_{i})$,
then the additivity properties of coherence are naturally required:
if we consider two (possibly subsequent) independent experiments $i$
and $i^{\prime}$ on the different non overlapping areas $D_{i}$ and
$D_{i^{\prime}}$, we would expect that the random quantity
$Y_{i}+Y_{i^{\prime}}$ stems from the same stochastic model with
parameters $\theta(D_{i}\cup D_{i^{\prime}})$. A compound Poisson
distribution is a sum of independent identically distributed random
variables in which the number of terms in the sum has a Poisson
distribution. Compound Poisson distributions are candidate models
purposely tailored to verify the previous desired infinite
divisibility property since the class of infinitely divisible
distributions coincides with the class of limit distributions of
compound Poisson distributions (\cite{Fellerv271}, theorem 3 of
chapter 27).

Depending on the nature of the term in the random sum, the compound
distribution can be discrete or continuous. The construction of such
a compound distribution with an exponential random mark for
continuous data and with a geometric one for counts is recalled in
section 2. This approach is worthwhile for two reasons. The first is
parsimony : there is only one parameter for the Poisson distribution
plus an additional one for the probability distribution function
-~\emph{pdf}~- of each component of the random sum. Secondly, the
compound construction may assist our understanding in cases where
the data collection can be interpreted in terms of sampling a latent
marked Poisson field. That is to say that the data appear in latent
"clumps" that are "harvested" during the experiment, the Poisson
parameter being the presence intensity of such clumps. A random
variable is used to mimic the quantity (or the number of individuals
in the discrete case) independently in each clump. At the upper
level of the hierarchy, random effects are added to depict
heterogeneous conditions between blocks of experiments.

In section 3, we develop a stochastic version of the EM algorithm
\citep{dempster+78} with a Monte-Carlo step (using importance
sampling) for this non Gaussian random effect model with
zero-inflated data. Maximum likelihood estimates and the
corresponding variance-covariance matrix are derived. The
computational task remains rather tractable thanks to simplifying
gamma-exponential conjugate properties in the continuous case (and
beta-geometric conjugacy in the discrete case).

In section 4, the hierarchical model \citep{Wickle+98} with compound
Poisson distribution for zero-inflated data is exemplified using a
real case study with two marine species, urchin and starfish
abundance data from a scientific bottom-trawl survey of the southern
Gulf of St. Lawrence, Canada. The EM algorithm performs well in
obtaining the maximum likelihood estimates of parameters, but for
one of the two species we notice some discrepancy between the actual
coverage of the confidence intervals and their theoretical levels
(as given by the asymptotic normal approximation). Consequently, we
further focus on variance covariance matrix estimation in section 5
and investigate via simulation the behavior of coverage level of
confidence intervals for various experimental designs, in search of
a practical fulfillment of the asymptotic conditions. Finally, we
briefly discuss some inferential and practical issues encountered
when implementing such hierarchical models for zero-inflated data.

\section{Model construction}

We propose a hierarchical construction to represent data with extra
zero collected over a non-homogeneous area. The model is divided
into two main layers~: in the first one, we model the sampling
process within a homogeneous sub-area (strata) and in the second
layer, we introduce heterogeneity between strata at the top of the
hierarchy using random effects. The first subsections detail the
hierarchical constructions for continuous data. In the last
subsection \ref{sec:discrete}, we sketch out an obvious modification
to represent count data.

\subsection{Compound Poisson process to introduce extra zeros}
\label{sec:extra0} Imagine that data $Y$ are obtained by harvesting
an area $D$ and that there are some clumps distributed according to
an homogeneous Poisson  process~: clumps are uniformly distributed
with a constant intensity, say  $\mu$.

\begin{figure}[h!]
  \centering{\scalebox{0.4}{\includegraphics*{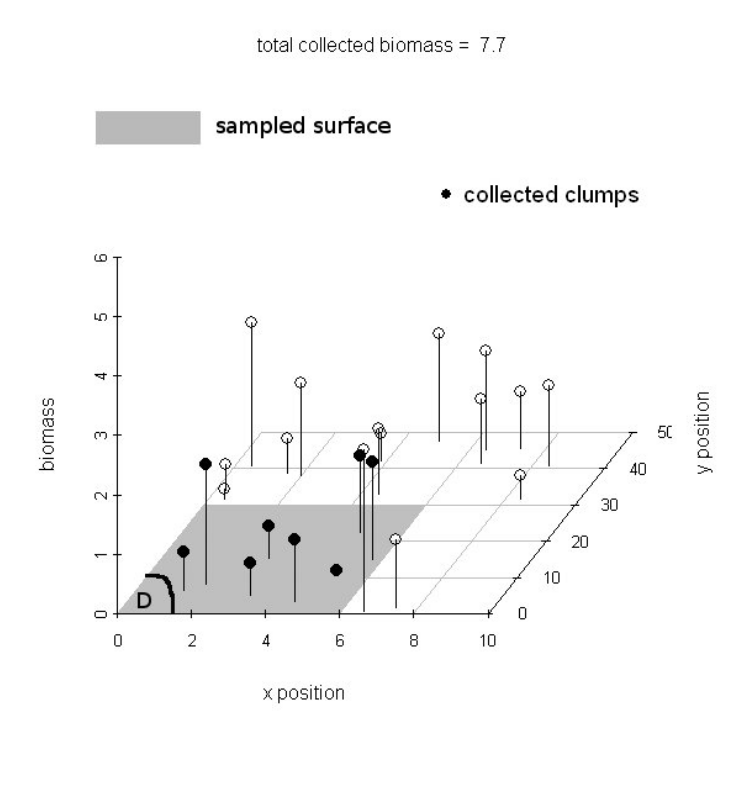}}}\caption[Realization of a compound Poisson distribution]{ Realization of a marked Poisson process on a region of $\mathbb{R}^{2}$, the sample is conducted over a region $D$. Here the total catch is $y=7.7$, the effective number of collected clumps is $8$.}
  \label{Fig:LolEx}
\end{figure}

By harvesting an area $D$, we pick an integer-valued random variable $N$ of clumps. According to Poisson process property $N$ follows a Poisson distribution of parameter $\mu\, D$.
For each clump $i$ the independent random variables $X_{i}$ or \emph{marks}
(with the same probability distribution) represent for instance the possible biomass in each  clump to be collected.\\
The final return will consist of the sum over $N$ clumps of the amount contained in each clump. With the convention that $Y=0$ if $N=0$ , the random sum~:
\begin{equation}
  \label{compounddef}
  Y=\sum_{i=0}^{N}X_{i},
 \end{equation}
is said to follow a compound Poisson distribution.
Figure~\ref{Fig:LolEx} exemplifies a realization of the total amount
of a collect
(i.e., sum of the marks) in a sampled region $D$.\\
\medskip

The Poisson-based additivity property avoids the drawback of
classical models mentioned in the introduction. Generally, $D$ is
the area of the sampled area included in $\mathbb{R}^{2}$. We assume
an homogeneous region $\mu(D)=\mu\, D$, so that the expected number
of collected clumps is proportional to the catching effort. The
difficulty with the generalization to an inhomogeneous Poisson
process lies in the inference step, not in the modeling step.
Consequently we used another approach to deal with heterogeneity
(see section \ref{sec:inhomogene}).
 In the following, we mostly omit to index quantities with this catching effort
 for presentation clarity, explicitly mentioning it only when necessary.

 Summary statistics about such compound distribution $Y$ are easily obtained (the
characteristic function is given in appendix \ref{an:fc})~:

 \begin{eqnarray*}
   \mathbb{E}(Y)  & =\mu\,D\ \mathbb{E}(X)\\
   \mathbb{V}ar(Y)  &  =\mu\,D\ \mathbb{E}(X^{2})
 \end{eqnarray*}

Parameter $\mu$ rules the occurrence of zero values when assuming $\P(X=0)=0$ \emph{i.e.} that the random mark is non atomic at $0$~:

$$\P(Y=0)=\exp{\left( -\mu\, D \right) }.$$

\subsection{Choice of the random component $X$ for continuous data}

For real-valued data with extra zeros, we will concentrate in this paper on
the exponential distribution of parameter $\rho$ for component $X$ \ such that
$\mathbb{E}(X)=\rho^{-1}$, leading to
\[
\mathbb{E}(Y)=\frac{\mu\,D}{\rho}\quad\text{and}\quad\mathbb{V}ar(Y)=2\frac
{\mu\,D}{\rho^{2}}.
\]

To keep on with an ecological interpretation of the model, assuming
that the mark $X$ follows an exponential distribution of parameter
$\rho$, means for the biologist that the probability of finding a
large amount of biomass within a clump is exponentially decreasing
and that the average quantity in each clump is $\rho^{-1}$. When no
clump is collected, there occurs a zero for the model $Y$. We choose
the exponential distribution because of parsimony and because of its
interesting conjugate property detailed in section
\ref{sec:expstep}.

This compound Poisson distribution was termed law of leaks (LOL) by
\cite{BernierFandeux70}, where $X$ represents elementary unobserved
leaks occurring at $N$ holes (uniformly located) along a gas
pipeline. In summary~:
\begin{equation}
\left(  Y\sim LOL(\mu,\rho)\right)  \Longleftrightarrow\left(
\begin{array}{l}
 Y=\sum_{i=1}^{N}X_{j},\cr
N\sim\mathcal{P}(\mu), \cr
(X_{1},\ldots,X_{N})\overset{i.i.d}{\sim}\mathcal{E}(\rho)
\end{array}
\right)  \label{eq:LOL}
\end{equation}

For the discrete case, a similar definition holds with the
corresponding geometric distribution for the marks (see section
\ref{sec:discrete}).

\subsection{Random effects}

\label{sec:inhomogene}

\begin{figure}[h!]
\centering{\scalebox{0.3}{\includegraphics*{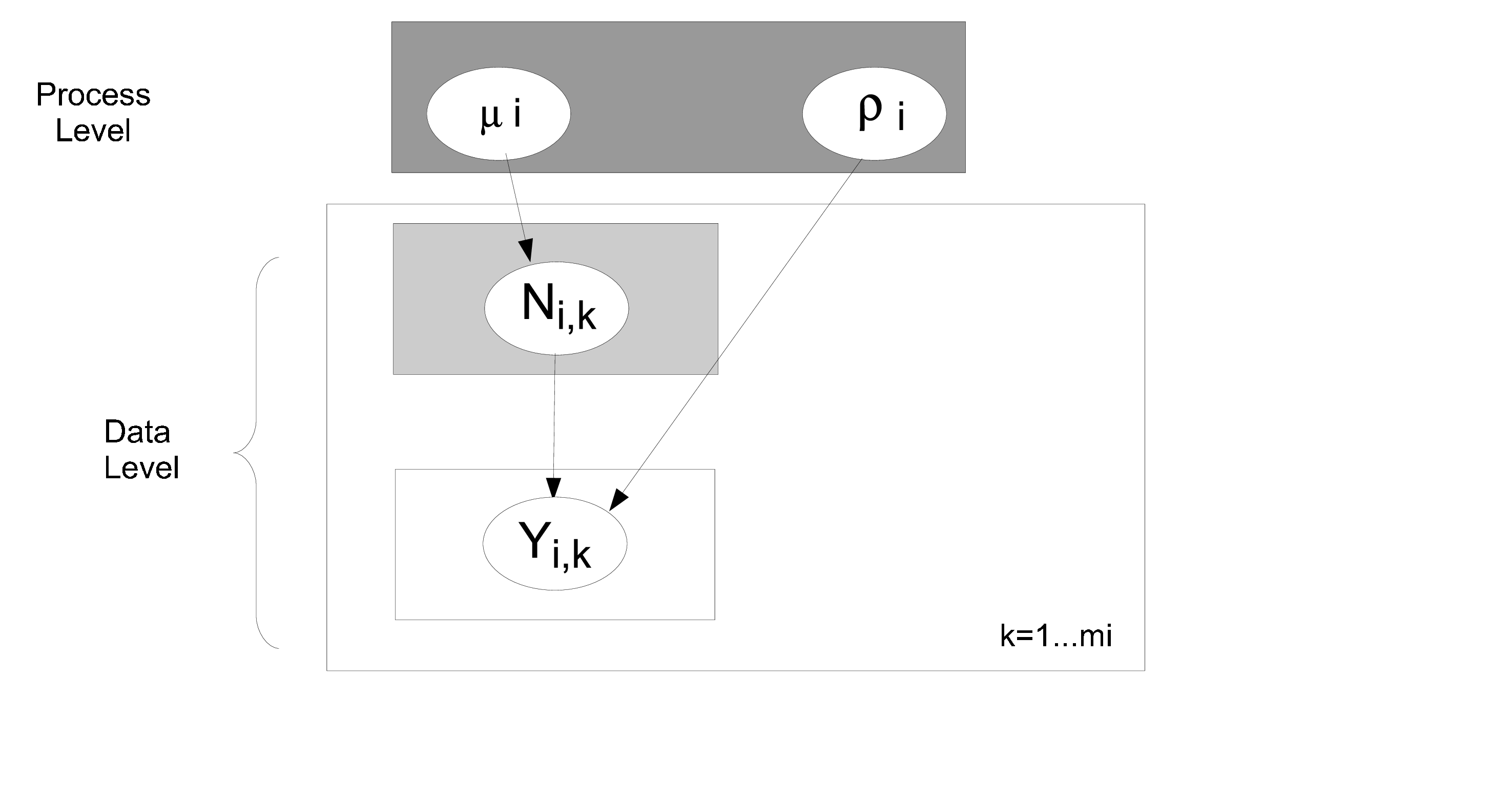}}}\caption{DAG of the
RLOL model}%
\label{Fig:DagRlol}%
\end{figure}

Although the previous compound construction could have formally been
extended to non-homogeneous Poisson processes, it is easier but
still quite realistic to relax the assumption of homogeneity by
considering homogeneous blocks (or strata), modeling possible
inter-block dispersion using random effects. We consider $S$
blocks~; in a given block $s$ there are $I_{s}$ grouped
observations. We denote by $\underline
{Y_{s}}=(Y_{s\,1},\ldots,Y_{s\,I_{s}})$ the random vector in block
$s$ and by
$\underline{\mathbf{Y}}=(\underline{Y_{1}},\ldots,\underline{Y_{S}})$
the whole vector over the $S$ blocks. The coefficients $a$ and $b$
of the gamma \emph{pdf} $\Gamma(a,b)$ for a random variable $\mu$
are such that $\mathbb{E}(\mu)=\frac{a}{b}$ and
$\mathbb{V}ar(\mu)=\frac{a}{b^{2}}$. The random effect model
$RLOL(a,b,c,d)$\ representing the occurrence of the sample
$\underline{\mathbf{Y}}$ is defined by the following set of
equations.
\begin{equation}
\underline{\mathbf{Y}}\sim RLOL(a,b,c,d)\Longleftrightarrow\left\{
\begin{array}
[c]{l}%
(\mu_{1},\ldots,\mu_{S})\overset{i.i.d}{\sim}\Gamma(a,b),\\
(\rho_{1},\ldots,\rho_{S})\overset{i.i.d}{\sim}\Gamma(c,d),\\
Y_{s,1},\ldots,Y_{s,I_{s}}\mid\mu_{s},\rho_{s}\overset{i}{\sim
}LOL\left( \mu_{s}\, D_{s,k} ,\rho_{s}\right)  \ \forall
s\in\left\{  1,\ldots,S\right\}.\end{array}
\right.  \label{eq:RLOL1}%
\end{equation}
The choice of a gamma distribution for the random effect is
motivated by conjugate properties which are useful in the inference
of the model. Section \ref{sec:gammaassumption} will show that it
may also be quite a realistic distribution for some datasets. The
hierarchical construction is summed up by the directed acyclic graph
(DAG as termed by \cite{Spie+96}) in Figure~\ref{Fig:DagRlol}.

\subsection{Compound Poisson process for count data}
\label{sec:discrete} A similar but discrete version to model count
data, can be obtained by changing the nature of the random marks of
the Poisson process. In this paper, we study a geometric
distribution with parameter $p=\P(X=1)$. The core of the model is
thus given by the following compound Poisson process with geometric
marks~:

$$
\left(  Y\sim DLOL(\mu,p)\right)  \Longleftrightarrow\left(
\begin{array}{l}
 Y=\sum_{i=1}^{N}X_{j},\cr
N\sim\mathcal{P}(\mu), \cr
(X_{1},\ldots,X_{N})\overset{i.i.d}{\sim}\mathcal{G}(p)
\end{array}
\right)
$$

To preserve conjugate properties, the gamma distribution for the random effect on the marks is replaced by a beta distribution so that the count data
version of the model is given by~:
\begin{equation}
\underline{\mathbf{Y}}\sim RDLOL(a,b,c,d)\Longleftrightarrow\left\{
\begin{array}
[c]{l}%
(\mu_{1},\ldots,\mu_{S})\overset{i.i.d}{\sim}\Gamma(a,b),\\
(p_{1},\ldots,p_{S})\overset{i.i.d}{\sim}\beta(c,d),\\
Y_{s,1},\ldots,Y_{s,I_{s}}\mid\mu_{s},p_{s}\overset{i}{\sim
}DLOL(\mu_{s} D_{s,k},p_{s})\ \forall s\in\left\{  1,\ldots,S\right\}  .
\end{array}
\right.  \label{eq:RDLOL1}%
\end{equation}
where $DLOL$ means Discrete version of Law of leaks and $RDLOL$
discrete law of leaks with random effects.

In most of the paper, we will simply state the main results when
technical aspects of the proofs are shared between discrete and
continuous cases.

\section{Estimation via the EM algorithm with importance sampling}
\label{sec:EM} Hierarchical models such as \ref{eq:RLOL1} or
\ref{eq:RDLOL1} cannot be straightforwardly estimated because of the
latent variables. The random effects $(\mathbf{\mu},\mathbf{\rho})$
and the unknown numbers of clumps $\underline {\mathbf{N}}$ must be
integrated out to obtain the likelihood. The likelihood has no
closed form and estimators cannot be directly derived. In such a
case, a classical strategy is to use Expectation Maximization
algorithm (\cite{dempster+78}) to derive max-likelihood estimates.
In our case the E step is not analytically accessible.   An
alternative is to use a stochastic version of this EM algorithm such
as Monte-Carlo EM ( MCEM see \cite{McCulloch94} or
\cite{McCulloch97}) or stochastic approximation of EM (SAEM see
\cite{Deylon99}).

 We detail in this section how to implement a MCEM algorithm using Importance sampling to obtain the maximum likelihood estimation and its
empirical variance matrix. Similar results concerning count data
process are summed up in the last subsection. From this point
onwards we will use brackets to denote \emph{pdf}'s as many
conditioning terms will appear in the probabilistic expressions
derived from the model fully specified by the set of equations
(\ref{eq:RLOL1}). The brackets denote either a density or a discrete
probability distribution, as in \cite{gelfand+90}. Following
Bayesian conventions, we will also allow the parameters to appear as
conditioning terms (\emph{i.e.}, instead of writing $\P(X)$ we will
specify $[X|a,b,c,d]$) so as to help the reader understand which
layer of the hierarchical model (\ref{eq:RLOL1}) the probability
expression refers to (see Fig \ref{Fig:DagRlol}).

\subsection{Implementation of the MCEM algorithm}

In this paper, $\theta$ stands for the set of parameters
$(a,b,c,d)$ in the $RLOL$ model. Given the random effects, the
data within a block are independent~:

$$L(\theta;\underline{\mathbf{Y}},\underline{\mathbf{N}},\mathbf{\mu
},\mathbf{\rho})=\sum_{s=1}^{S}L_{s}$$

where $L_{s}$ denotes the \emph{complete} log-likelihood in block
$s$, i.e.~:

\begin{align}
L_{s}=L_{s}(\theta;\underline{Y_{s}},\underline{N_{s}},\mu_{s},\rho_{s})  &
=\left(  \sum_{i=1^{I_{s}}}\ln\left(  \left[  Y_{s,i}|N_{s,i},\rho_{s}\right]
\left[  N_{s,i}|\mu_{s}\right]  \right)  \right)
+\label{blockcompleteloglikelihood}\\
&  \hspace{1cm}\ln\left(  \left[  \mu_{s}|a,b\right]  \right)  +\ln\left(
\left[  \rho_{s}|c,d\right]  \right) \nonumber
\end{align}

Following \cite{Tanner1996}, the pivotal quantity in the EM
algorithm (recalled in appendix \ref{an:EM}) is the conditional
expectation of the complete log-likelihood~:

$$
Q(\theta,\theta^{\prime})=\mathbb{E}_{\theta^{\prime}}\left(  L(\theta
;\underline{\mathbf{Y}},\underline{\mathbf{N}},\mathbf{\mu},\mathbf{\rho
})|\underline{\mathbf{Y}}\right)
$$

\subsubsection{Maximization step}

To maximize $Q(\theta,\theta^{\prime})$ with respect to $\theta $,
we focus on the terms that involve $\theta$~:

{\small
\begin{align}
Q(\theta,\theta^{\prime})  &  =C_{-\theta}(Y)+(a-1)\times\sum_{s=1}%
^{S}\mathbb{E}_{\theta^{\prime}}\left(  \ln{\mu_{s}}\mid\underline{Y_{s}%
}\right)  +Sa\ln{b}-b\sum_{s=1}^{S}\mathbb{E}_{\theta^{\prime}}\left(  \mu
_{s}\mid\underline{Y_{s}}\right)  -S\ln(\Gamma(a))\nonumber\label{eq:QRLOL}\\
&  +(c-1)\times\sum_{s=1}^{S}\mathbb{E}_{\theta^{\prime}}\left(  \ln{\rho_{s}%
}\mid\underline{Y_{s}}\right)  +Sc\ln{d}-d\sum_{s=1}^{S}\mathbb{E}%
_{\theta^{\prime}}\left(  \rho_{s}\mid\underline{Y_{s}}\right)  -S\ln
(\Gamma(c)),
\end{align}
} where $C_{-\theta}(Y)$ denotes a constant which does not depend on
$\theta$.

Differentiating with respect to $\theta$, we obtain the set of
equations to be satisfied at the maximum\ $\underset{\theta}{argmax\
}Q(\theta,\theta^{\prime })$:

\begin{equation}
\label{eq:max1}
\frac{a}{b}=\frac{
{\displaystyle\sum\limits_{s=1}^{S}}
\mathbb{E}_{\theta^{\prime}}\left(  \mu_{s}\mid\underline{Y_{s}}\right)  }{S}
\end{equation}

\begin{equation}
\label{eq:max2}
\ln{a}-\psi(a)=\ln{\left(  \frac{
{\displaystyle\sum\limits_{s=1}^{S}}
\mathbb{E}_{\theta^{\prime}}\left(  \mu_{s}\mid\underline{Y_{s}}\right)  }%
{S}\right)  }-\frac{{\displaystyle\sum\limits_{s=1}^{S}}
\mathbb{E}_{\theta^{\prime}}\left(  \ln{\mu_{s}}\mid\underline{Y_{s}}\right)
}{S}
\end{equation}

\begin{equation}
\frac{c}{d}=\frac
{\displaystyle\sum\limits_{s=1}^{S}
\mathbb{E}_{\theta^{\prime}}\left(  \rho_{s}\mid\underline{Y_{s}}\right)  }{S}
\label{eq:max3}
\end{equation}

\begin{equation}
\ln{c}-\psi(c)=\ln{\left(  \frac{
{\displaystyle\sum\limits_{s=1}^{S}}
\mathbb{E}_{\theta^{\prime}}\left(  \rho_{s}\mid\underline{Y_{s}}\right)  }%
{S}\right)  }-\frac{
{\displaystyle\sum\limits_{s=1}^{S}}
\mathbb{E}_{\theta^{\prime}}\left(  \ln{\rho_{s}}\mid\underline{Y_{s}}\right)
}{S} \label{eq:max4}
\end{equation}

$\psi(x)$ denotes the digamma function defined as the first
logarithmic derivative of $\Gamma(x)$. No analytical expression
can be derived for $\theta$ as the argument of the maximum of
$Q(\theta,\theta^{\prime})$, but a Newton-Raphson algorithm
\medskip is efficient and easy to implement with a good empirical starting
point as indicated in annex \ref{an:NR}.\

\subsubsection{Expectation step by conditioning onto the number of clumps}

\label{sec:expstep}

The right-hand side of equations \ref{eq:max1} to \ref{eq:max4} involves
$\mathbb{E}_{\theta^{\prime}}\left(  \mu_{s}\mid\underline{Y_{s}}\right)  $,
$\mathbb{E}_{\theta^{\prime}}\left(  \ln(\mu_{s})\mid\underline{Y_{s}}\right)
$, $\mathbb{E}_{\theta^{\prime}}\left(  \rho_{s}\mid\underline{Y_{s}}\right)
$ and $\mathbb{E}_{\theta^{\prime}}\left(  \ln(\rho_{s})\mid\underline{Y_{s}%
}\right).$ To compute these expected values, we will proceed by conditioning
onto the hidden number of clumps $\underline{\mathbf{N}}$. Proposition
\ref{prop:murhocond} shows that, given $\underline{\mathbf{N}},$ these four
target quantities are simply marginal expectations of the sufficient quantity
$N_{s+},$ the only necessary function of $\underline{\mathbf{N}}$ \ that needs
to be evaluated within each block $s$.\par

In a second step, integration over the number of clumps is performed
by recourse to importance sampling within a block $s$ as detailed in
proposition \ref{prop:IS}. Proofs of propositions are given in
appendix \ref{subsec:proof}

\begin{prop}
\label{prop:murhocond} Assuming $Y\sim RLOL(\theta')$ with
$\theta'=(a',b',c',d')$, $S$ strata   and $I_{s}$ records in stratum
$s$ as in \ref{eq:RLOL1} , then the complete conditional
distributions of $\mu_{s}$ and $\rho_{s}$ in one particular stratum
$s$ are given by

\begin{equation}
\mu_{s}|\underline{\mathbf{N}},\underline{\mathbf{Y}},\theta'\sim
\Gamma(a'+N_{s+},b'+D_{s+}),
\end{equation}

and

\begin{equation}
\rho_{s}|\underline{\mathbf{N}},\underline{\mathbf{Y}},\theta'\sim
\Gamma(a'+N_{s+},b'+Y_{s+}),
\end{equation}

where in stratum $s$, $N_{s+}=\sum_{i=1}^{I_{s}}N_{si}$ denotes the
total number of clumps caught, $Y_{s+}=\sum_{i=1}^{I_{s}}Y_{si}$ is the entire quantity harvested and $D_{s+}=\sum_{i=1}^{I_{s}}D_{si}$ is the whole catching effort.\\
The quantities involved in the E step are given by~

\begin{align}
\mathbb{E}_{\theta'}\left(  \mu_{s}\mid\underline{y_{s}}\right)  =  &
\frac{a'+\mathbb{E}_{\theta'}\left(  N_{s+}|\underline{y_{s}}\right)  }%
{b'+D_{s+}},\label{eq:N+1}\\
\mathbb{E}_{\theta'}\left(  \ln(\mu_{s})\mid\underline{y_{s}}\right)  =  &
\mathbb{E}_{\theta'}\left(  \psi(a'+N_{s+})\left|  \underline{y_{s}}\right.
\right)  -\ln(b'+D_{s+}),\label{eq:N+2}\\
\mathbb{E}_{\theta'}\left(  \rho_{s}\mid\underline{y_{s}}\right)  =  &
\frac{c'+\mathbb{E}_{\theta'}\left(  N_{s+}|\underline{y_{s}}\right)  }%
{d'+Y_{s+}},\label{eq:N+3}\\
\mathbb{E}_{\theta'}\left(  \ln(\rho_{s})\mid\underline{y_{s}}\right)  =  &
\mathbb{E}_{\theta'}\left(  \psi(c'+N_{s+})\left|  \underline{y_{s}}\right.
\right)  -\ln(d'+Y_{s+}). \label{eq:N+4}%
\end{align}
\end{prop}

This result merely comes from the conjugacy property between gamma and
Poisson distributions for $\mu$ (gamma and exponential distribution
concerning $\rho$). The moments of gamma and log gamma, beta and
log beta distributions are recalled in appendix \ref{an:Moments}.\par

In order to go one step further into the calculus, we have to
perform the integration over $N_{+}$. Proposition \ref{prop:Ncond}
gives the distribution of $N_{+}|Y_{+},\theta$ up to a constant.
Subsequently, the integration over $N_{+}$ will make recourse to
importance sampling as proposed in \cite{Levine2001}. This Monte
Carlo algorithm is detailed in proposition \ref{prop:IS}.

\begin{prop}
\label{prop:Ncond} Assuming $Y\sim RLOL(a,b,c,d)$ with $S$ strata, and $I_{s}$
records in stratum $s$, the conditional distribution of $\underline{N_{s}%
}|\theta,\underline{y_{s}}$ is given (up to a constant $K$) by
\begin{equation}
\lbrack\underline{N_{s}}|\theta',\underline{y_{s}}]=K\,\left(  \frac
{\Gamma(a'+N_{s+})\Gamma(c'+N_{s+})}{b'+D_{s+})^{N_{s+}}(d'+Y_{s+})^{N_{s+}}%
}\right)  \,\prod_{{\small i=1,\,y_{i}>0}}^{I_{s}}\left(  \frac{y_{si}%
^{N_{si}}}{\Gamma(N_{si})\Gamma(N_{si}+1)}\right)  \prod_{{\small i=1,\,y_{i}=0}}^{I_{s}}\delta(N_{si}) \label{eq:margeN}%
\end{equation}
\end{prop}

To draw a sample according to the rather intricate looking distribution \ref{eq:margeN}, an importance sampling based algorithm is detailed in the following proposition for one replicate (often termed \emph{particle)}. In order to obtain a $G$-sample, this procedure is repeated for each block $G$ times.

\begin{prop}
[Generate one particle in one particular stratum $s$ according to distribution
\ref{eq:margeN}]\label{prop:IS} A \emph{particle} $g$ is a vector
$(N_{s+}^{(g)},N_{s1}^{(g)},\ldots,N_{sI_{s}}^{(g)})$ in a particular stratum
$s$. Omitting $s$ to make the reading easier, we may assume with no loss of generality that the first $I^+$ terms are non zero and the $I-I^+$ followings are the zero ones. The algorithm to generate one \emph{particle} $g$ runs as follows:

\begin{enumerate}
\item  Generate $N_{i}^{(g)}=0$ wherever $y_{i=0}$ for
$i=I-I^{+}+1,\ldots,I$.

\item  Generate the\ value of the random sum $N_{+}^{(g)}$ according to the
importance distribution~: {\small
\[
f_{IS}(N^{+})\varpropto\left(  \frac{1}{b^{\prime}+D_{+}}\right)  ^{N^{+}%
}\left(  \frac{Y^{+}}{d^{\prime}+Y^{+}}\right)  ^{N^{+}}\frac{\Gamma
(a^{\prime}+N^{+})\Gamma(c^{\prime}+N^{+})}{\left(  \prod_{i=1}^{I^{+}}%
\Gamma\left(  \frac{y_{i}}{Y^{+}}N^{+}+1\right)  \right)  \Gamma\left(
N_{+}-I_{+}+1\right)  }%
\]
}

As the one dimensional importance distribution\ $f_{IS}$ is a quickly
decreasing function of $N^{+}$, its normalizing constant can be easily
approximated and a bounded interval is used in practice as the support of
$N^{+}$.

\item  Generate each $N_{i}^{(g)}$ for $i=1,\ldots,I^{+}$ so that the vector
$(N_{1}^{(g)}-1,\ldots,N_{I^{+}}^{(g)}-1)$ is distributed according to a
multinomial distribution $\mathcal{M}(N_{+}^{(g)}-I_{+},(y_{1}/Y_{+}%
,\ldots,y_{I^{+}}/Y_{+}))$.

\item  Associate to the vector $(N_{+}^{(g)},N_{1}^{(g)},\ldots,N_{I}^{(g)})$
generated at the previous step, the importance weight~:
\[
w^{(g)}=\prod_{i=1}^{I_{+}}\frac{\Gamma\left(  N_{+}^{(g)}\frac{y_{i}}{Y_{+}%
}+1\right)  }{\Gamma(N_{i}^{(g)}+1)}%
\]
\end{enumerate}
\end{prop}

The proof of this proposition is straightforward from importance
sampling theory (see for instance chapter 3 of
\cite{Robert+98}).\newline

The weighted sample of $N_{+}$ may be used to approximate the
expected conditional value defined in equations \ref{eq:N+1} to
\ref{eq:N+4}. For instance, quantity \ref{eq:N+2} is approximated
by~:
\[
\mathbb{E}_{\theta^{\prime}}\left(  \ln(\mu_{s})\mid\underline{y_{s}}\right)
\approx\left(  \frac{1}{\sum_{g=1}^{G}\omega^{(g)}}\sum_{g=1}^{G}\omega
^{(g)}\times\psi(a^{\prime}+N_{s+}^{(g)})\right)  -\ln(b^{\prime}+D_{s+}).
\]

\subsubsection{Empirical Variance Matrix}

This section is devoted to the evaluation of the empirical variance
matrix, so as to provide confidence regions. Because of the EM
principle, we assume that the algorithm has converged to the maximum
likelihood value $\hat{\theta}$. The empirical Fisher information
matrix is then given by proposition \ref{prop:Fisher}. To explicitly
compute this information matrix, we propose to numerically integrate
over $\underline{\mathbf{N}}$ thanks to importance sampling as
performed for the point estimation step. Technical details are also
given in appendix \ref{an:secondderiv}.

\begin{prop}
\label{prop:Fisher} Assuming $Y\sim RLOL(a,b,c,d)$ with $\ S$
strata, and $I_{s}$ records in stratum $s$ as in \ref{eq:RLOL1}. Let
us denote $I_{e}(\theta)$ the empirical information matrix defined
by
\begin{equation}
I_{e}(\theta)=-\frac{\partial^{2}\ln{[\underline{\mathbf{Y}}|\theta]}%
}{\partial\theta_{i}\,\partial\theta_{j}}%
\end{equation}
At the maximum likelihood estimator $\hat{\theta}$, the following
equality holds~:
\begin{equation}
I_{e}(\hat{\theta},\underline{\mathbf{Y}})=S\,\left(
\begin{matrix}
-\psi^{\prime}(\hat{a}) & \frac{1}{\hat{b}} & 0 & 0\\
\frac{1}{\hat{b}} & -\frac{\hat{a}}{\hat{b}^{2}} & 0 & 0\\
0 & 0 & -\psi^{\prime}(\hat{c}) & \frac{1}{\hat{d}}\\
0 & 0 & \frac{1}{\hat{d}} & -\frac{\hat{c}}{\hat{d}^{2}}%
\end{matrix}
\right)  +\sum_{s=1}^{S}(A_{s}+B_{s})
\end{equation}
with

$$
A_{s}=\left(
\begin{matrix}
\mathbb{E}_{\nu_{s}}(\psi^{\prime}(\etoile{a_{s}})) & \frac{-1}{\etoile{b_{s}}} & 0 & 0\\
\frac{-1}{\etoile{b_{s}}} & \frac{  \mathbb{E}_{\nu_{s}}(\etoile{a_{s}})}{  (\etoile{b_{s}})^{2}}
 & 0 & 0\\
0 & 0 & \mathbb{E}_{\nu_{s}}(\psi^{\prime
}(\etoile{c_{s}})) & \frac{-1}{\etoile{d_{s}}}\\
0 & 0 & \frac{-1}{\etoile{d_{s}}} & \frac{\mathbb{E}_{\nu_s}(\etoile{c_{s}})}{(\etoile{d_{s}})^{2}}
\end{matrix}
\right)
$$

and

$$
B_{s}=\left(
\begin{matrix}
\Var_{\nu_s}(\psi(\etoile{a_{s}})) & -\frac{\Cov_{\nu_s}(\etoile{a_{s}},\psi(\etoile{a_{s}}))}{\etoile{b_s}} & \Cov_{\nu_s}(\psi(\etoile{a_{s}}),\psi(\etoile{c_s})) & -\frac{\Cov_{\nu_s}(\etoile{c_s},\psi(\etoile{a_{s}}))}{\etoile{d_s}}\\
-\frac{\Cov_{\nu_s}(\etoile{a_s},\psi(\etoile{a_{s}})}{\etoile{b_s}} & \frac{\Var_{\nu_s}(\etoile{a_{s}})}{{\etoile{b_s}}^{2}} & -\frac{\Cov_{\nu_s}(\etoile{a_{s}},\psi(\etoile{c_s})}{\etoile{b_s}}& \frac{\Cov_{\nu_s}(\etoile{a_{s}},\etoile{c_s})}{\etoile{b_s}\,\etoile{d_s}}\\
\Cov_{\nu_s}(\psi(\etoile{a_s}),\psi(\etoile{c_s})) & -\frac{\Cov_{\nu_s}(\etoile{a_{s}},\psi(\etoile{c_s}))}{\etoile{b_s}}& \Var_{\nu_s}(\psi(\etoile{c_s})) & -\frac{\Cov_{\nu_s}(\etoile{c_s},\psi(\etoile{c_s}))}{\etoile{d_s}}\\
-\frac{\Cov_{\nu_s}(\etoile{c_{s}},\psi(\etoile{a_{s}})}{\etoile{d_s}} & \frac{\Cov_{\nu_s}(\etoile{a_{s}},\etoile{c_{s}})}{\etoile{b_s}\,\etoile{d_s}} & -\frac{\Cov_{\nu_s}(\etoile{c_s},\psi(\etoile{c_s})}{\etoile{d_s}}& \frac{\Var_{\nu_s}(\etoile{c_s})}{{\etoile{d_s}}^{2}}
\end{matrix}
\right),
$$

where $\etoile{a_s}=\hat{a}+N_{s+}$, $\etoile{b_s}=\hat{b}+D_{s+}$,
$\etoile{c_s}=\hat{c}+N_{s+}$, $\etoile{d_s}=\hat{d}+Y_{s+}$ and $\nu_s$ stands for the probability measure of $N_{s+}\vert \hat{\theta},\mathbf{\underline{Y}}$.
\end{prop}

As for the first derivative phase of the EM algorithm detailed in section
\ref{prop:IS}, the operations $\mathbb{E}_{\underline{\mathbf{N}}\vert \hat{\theta},\underline{\mathbf{Y}}}$ and $\Var_{\underline{\mathbf{N}}\vert \hat{\theta},\underline{\mathbf{Y}}}$ , needed to evaluate $A_{s}$ and
$B_{s}$, can be easily implemented by recourse to the very
same Monte-Carlo $N_{+}$ sample that was previously drawn by
importance sampling.

\subsubsection{Prediction of the random effects}
It is of interest to predict the random effects in each stratum, for
instance to help illustrate the heterogeneity between units. In a
linear mixed model context, the \emph{Best Linear Unbiased
Estimator} is defined by the conditional expectation of the random
effect according to the data $\mathbf{\underline{y}}$ and the point
estimation. We follow the same avenue of thought and define a
predictor of the random effects by the conditional expectation.
Using formula \ref{eq:N+1} and \ref{eq:N+3}, the random effect
predictors are given by~:

\begin{equation}\label{eq:murandeffpred}
\mu_s^{(pred)}=\E(\mu_s\vert \mathbf{\underline{y}},\hat{\theta})=\frac{\hat{a}+\mathbb{E}\left(  N_{s+}|\underline{y_{s}}, \hat{\theta} \right)  }%
{\hat{b}+D_{s+}},
\end{equation}

and

\begin{equation}
\rho_s^{(pred)}=\E(\rho_s\vert \mathbf{\underline{y}},\hat{\theta})=\frac{\hat{c}+\mathbb{E}\left(  N_{s+}|\underline{y_{s}}, \hat{\theta} \right)  }%
{\hat{d}+Y_{s+}},
\end{equation}


The following section aims at highlighting the differences between the continuous case detailed previously and the discrete one.
\subsection{MCEM algorithm for RDLOL model}

\subsubsection{Straightforward transposition to  the discrete case}

The definition of the model designed for the discrete case and called RDLOL model is given by equation \ref{eq:RDLOL1}, in this
case the pivotal quantity $Q(\theta,\theta^{\prime})$ reads~:

{\footnotesize
\begin{align}
Q(\theta,\theta^{\prime})  &  =C_{-\theta}(Y)+(a-1)\sum_{s=1}^{S}%
\mathbb{E}_{\theta^{\prime}}\left(  \ln{\mu_{s}}\mid\underline{Y_{s}}\right)
+Sa\ln{b}-b\sum_{s=1}^{S}\mathbb{E}_{\theta^{\prime}}\left(  \mu_{s}%
\mid\underline{Y_{s}}\right)  -S\ln(\Gamma(a))\nonumber\label{eq:QRDLOL}\\
&  +\ln\left(  \frac{\Gamma(c+d)}{\Gamma(c)\Gamma(d)}\right)  +(c-1)\sum
_{s=1}^{S}\mathbb{E}_{\theta^{\prime}}\left(  \ln{p_{s}}\mid\underline{Y_{s}%
}\right)  +(d-1)\sum_{s=1}^{S}\mathbb{E}_{\theta^{\prime}}\left(  \ln
(1-p_{s})\mid\underline{Y_{s}}\right)
\end{align}
}

The equations satisfied at the maximum \ for $(a,b)$ are again \ref{eq:max1} and \ref{eq:max2}$.$ Due to the substitution of a gamma \emph{pdf} into a beta \emph{pdf} for the random effects governing the geometric discrete marks in the random sum of counts, parameters $c$ and $d$ verify equations \ref{eq:maxDLOL1} and \ref{eq:maxDLOL2} (equivalent to equations \ref{eq:max3} and \ref{eq:max4} in the continuous data model) ~:
\begin{equation}
\label{eq:maxDLOL1}
\displaystyle\psi(c+d)-\psi(c) = -\displaystyle\frac{\displaystyle\sum
_{s=1}^{S} \displaystyle\mathbb{E}_{\theta^{\prime}}\left(  \ln p_{s}
\mid\underline{Y_{s}}\right)  }{S}%
\end{equation}

\begin{equation}
\label{eq:maxDLOL2}
\psi(c)-\psi(d) = \displaystyle\frac{\displaystyle\sum_{s=1}^{S}
\displaystyle\mathbb{E}_{\theta^{\prime}}\left(  \left.  \ln\left(
\frac{p_{s}}{1-p_{s}}\right)  \right\vert \underline{Y_{s}}\right)  }{S}%
\end{equation}

The approach used for the continuous case is reproduced to obtain, in each
stratum s the conjugate conditional density of $\underline{\mu}$, $\underline
{p}$~, so that the analog to propositions \ref{prop:murhocond} and \ref{prop:Ncond} is~:
\begin{prop}
\label{prop:mupcond}  Assuming $Y\sim RDLOL(\theta')$ with
$\theta'=(a',b',c',d')$, $S$ strata   and $I_{s}$ records in stratum
$s$ as in \ref{eq:RDLOL1} , then the complete conditional
distributions of $\mu_{s}$ and $p_{s}$ in one particular stratum $s$
are given by {\small
\begin{equation}
\mu_s|N_{s+},\theta^{\prime}\sim
\Gamma(a^{\prime}+N_{s+},b^{\prime}+D_{s+}),
\end{equation}
and
\begin{equation}
p_{s}|N_{s+},\theta^{\prime}\sim\beta(c^{\prime}+N_{s+},d^{\prime}%
+Y_{s+}-N_{s+}).
\end{equation}
}

Furthermore the conditional distribution function of $\underline{N_s}$ is~:
{\footnotesize
\begin{equation}
\left[  \underline{N_s}\vert  \theta',\mathbf{\underline{Y}}\right]  \propto\left(  \prod_{i=1}^{I^{+}}
  \frac{
    \left(
      \begin{array}{c}
        Y_{si}-1\\
        N_{si}-1\\
      \end{array}
    \right) D_{si}^{N_{si}}}
  {N_{si}!}\right)
\left(\prod_{i^=I-I^++1}^{I}\delta(N_{si})\right)
\left(  \frac{\Gamma(a^{\prime}+N_{s+})\Gamma(N_{s+}+c^{\prime})\Gamma
    (Y_{+}-N_{s+}+d^{\prime})}{(b^{\prime}+D_{s+})^{N_{s+}}}\right)
\label{eq:condNdiscretecase}
\end{equation}
}
\end{prop}

The choice of an efficient importance sampling distribution in the
discrete case is not the straightforward adaptation of the
continuous gives and a mixture has to be used to obtain an efficient
and well behaved algorithm, detailed in appendix \ref{app:DiscAlgo}.

\subsubsection{The covariance matrix in the discrete case}

The covariance matrix in the discrete case benefits from the same
conditional independence decompositions and the adaptation of the
continuous case is straightforward given the moments of the beta
distribution in appendix \ref{an:Moments}; the result is detailed in
appendix \ref{an:secondderivdiscrete}.  The weighted sample of
$N_{+}$ is used to compute the expectations and variance-covariance
terms in the matrix components.

\subsubsection{Prediction of the random effects}
The predictions of the random effects are just given by the
conditional expectations. Unsurprisingly, the predictions in the
discrete case and in the continuous one look very similar.
$\mu_s^{(pred)}$ is still given by formula \ref{eq:murandeffpred}
and

\begin{equation}
p_s^{(pred)}=\E(p_s\vert \mathbf{\underline{y}},\hat{\theta})=\frac{\hat{c}+\mathbb{E}\left(  N_{s+}|\underline{y_{s}}, \hat{\theta} \right)  }%
{\hat{c}+\hat{d}+Y_{s+}},
\end{equation}

\section{Applications}
In this section, we apply the EM estimation procedure to two real
datasets of ecological interest. We then study the validity of
asymptotic assumptions by assessing the coverage level of confidence
regions.

\subsection{Real dataset -  Gulf of St.Lawrence survey}
\label{sec:realdata} A multi-species bottom-trawl survey of the
southern Gulf of St.Lawrence (NW Atlantic) has been conducted each
September since 1971. The purpose of this survey is to estimate the
abundance and characterize the geographic distribution of marine
biota. The survey follows a stratified random design, with $38$
strata defined largely as homogeneous habitats using depth,
temperature and sediments properties. The target fishing procedure
at each fishing station is a 30-min straight-line tow at a speed of
3.5 knots (i.e., 3.21km trawled distance). However the actual
distance trawled can vary due to winds, currents and the avoidance
of damaging rough bottoms; sampling effort is therefore variable
among trawl tows, but this source of additional variability is
easily accommodated in the models presented here ( the $D_{s,k}$ in
eq \ref{eq:RLOL1}). For our case study, we use data on the abundance
of sea urchins and Sunflower starfishes collected during three
survey years (1999-2001), in a total of 540 bottom-trawl sets. The
time period was chosen to minimize inter-annual changes in abundance
while ensuring a sufficient sample size. The species were selected
because inter-annual changes in their geographic distribution
resulting from movements of individuals at the scale of survey
sampling can be assumed to be approximately nil.

\begin{figure}[h!]
  \begin{minipage}[b]{0.45\linewidth}
    \centering \includegraphics[scale=0.5]{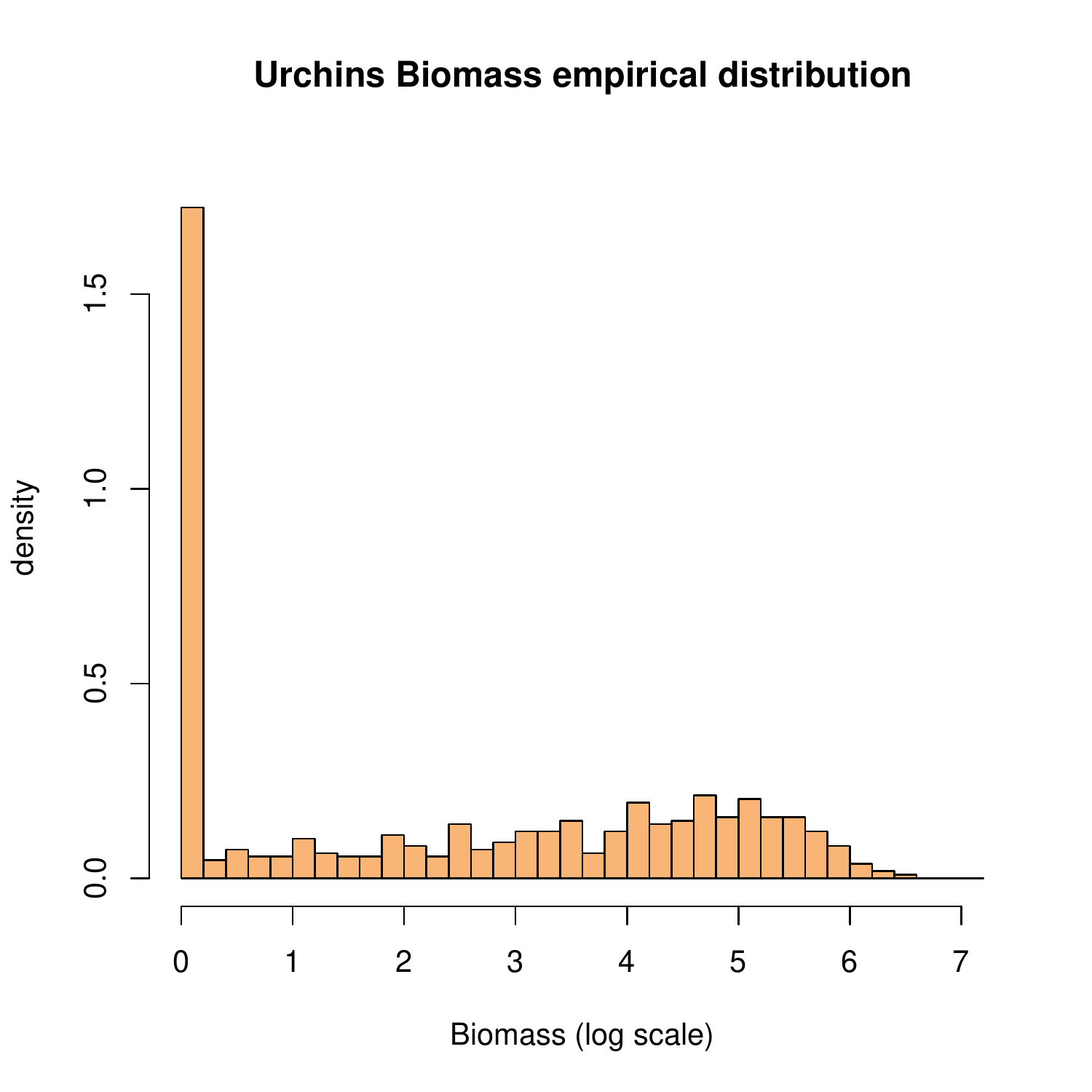}
    \caption{Histogram of urchin biomass (kg/tow) from individual tows in the
      southern Gulf of St. Lawrence, bottom-trawl surveys: 1999-2000-2001 }
    \label{Fig:Urchinhist}
  \end{minipage}\hfill
  \begin{minipage}[b]{0.45\linewidth}
    \centering \includegraphics[scale=0.5]{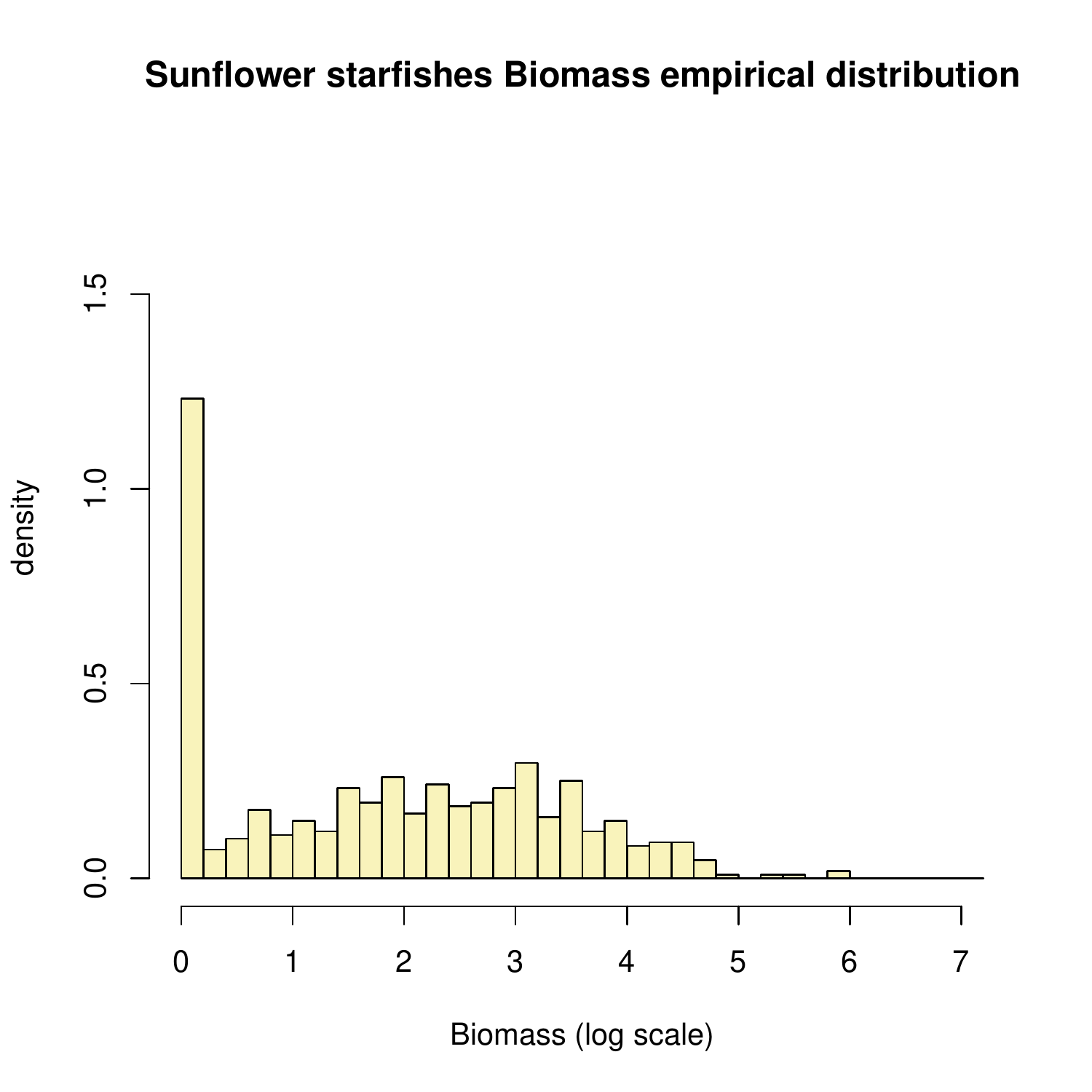}
    \caption{Histogram of Sunflower Starfishes biomass (kg/tow) from individual tows in the
      southern Gulf of St. Lawrence, bottom-trawl surveys: 1999-2000-2001 }
    \label{Fig:SunStarhist}
  \end{minipage}
\end{figure}

\begin{figure}[h!]
\begin{center}
\includegraphics[width=10 cm]{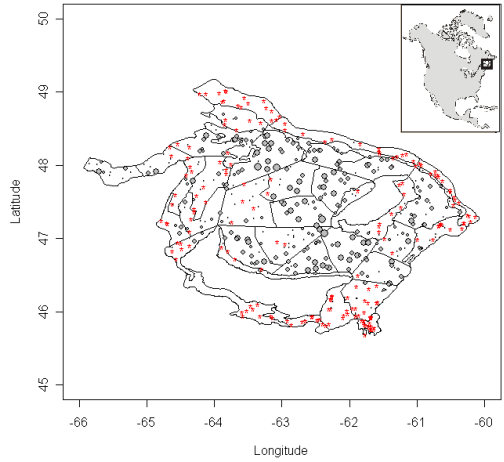}
\end{center}
\caption{Locations of urchin catches (symbols) and stratum boundaries
(lines) in the southern Gulf of St. Lawrence bottom-trawl surveys
1999-2000-2001. The radii of the circles are proportional to the
biomass (in kg/tow) caught. The "*" denote sites with
no urchins caught. Starfishes are not plotted.}%
\label{Fig:Urchindata}%
\end{figure}

The histograms of urchin and starfish catches in kg per survey tow
clearly reflect zero-inflated distributions
(Fig~\ref{Fig:Urchinhist} and \ref{Fig:SunStarhist}). A large number
of tows capture no urchin (nor starfish) and catches in non-zero
tows tend to follow a skewed distribution. At the scale of the
survey, sea urchins are distributed in patches of localized variable
abundance, interspersed by numerous and relatively large areas where
the species is absent (Fig \ref{Fig:Urchindata}). Such \emph{patchy}
distributions of organisms are prevalent in ecological science. Data
in two strata are always zero, thus rendering estimation impossible
if we were to fit one model per stratum or to consider $\rho_s$ as
fixed effects. Because the hierarchical framework allows some
transfer of information between strata, the other data help to
predict $\rho$ in these two strata.

\subsubsection{Maximum likelihood point estimation}
The estimation procedure follows the EM algorithm detailed in
appendix \ref{an:EM} (with a stopping rule when the sixth decimal
does not change between iterations) and gives  values of
\begin{equation*}
    \hat{\theta}^{Urch}=(\hat{a},\hat{b},\hat{c},\hat{d})=(0.997797,1.05107,5.05733,13.0312),
\end{equation*}
and
\begin{equation*}
    \hat{\theta}^{Sun}=(\hat{a},\hat{b},\hat{c},\hat{d})=(1.91879,1.80704,1.90002,0.898734),
\end{equation*}

as a maximum likelihood point estimates respectively for Urchin
 and Sunflower starfishes datasets.

\begin{figure}[h!]
  \begin{minipage}[b]{0.45\linewidth}
    \centering \includegraphics[scale=0.5]{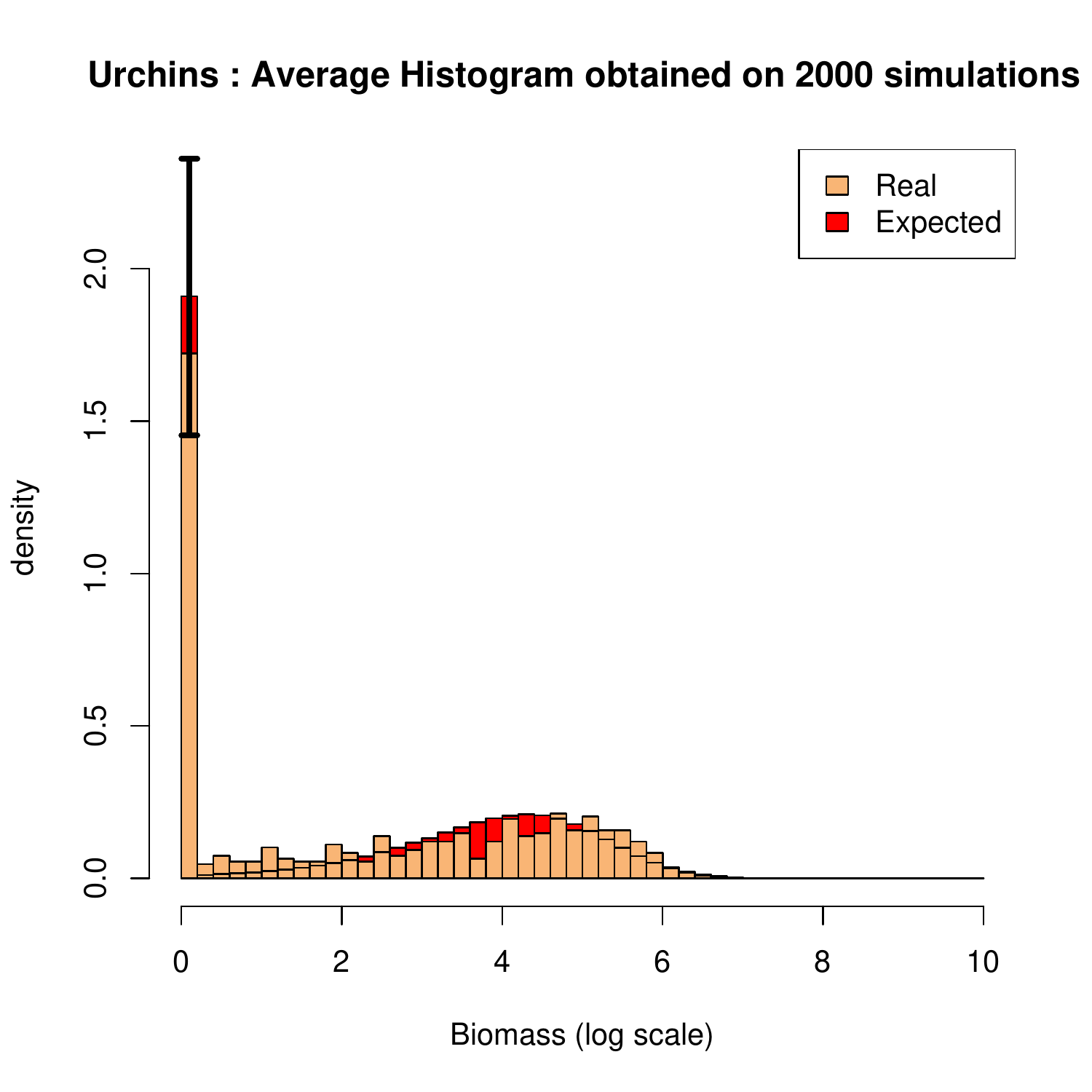}
    \caption{Comparisons between  urchins dataset and averaged histogram (1000 simulations of datasets at $\hat{\theta}^{Urch}$)}
    \label{fig:urchinfit}
  \end{minipage}\hfill
  \begin{minipage}[b]{0.45\linewidth}
    \centering \includegraphics[scale=0.5]{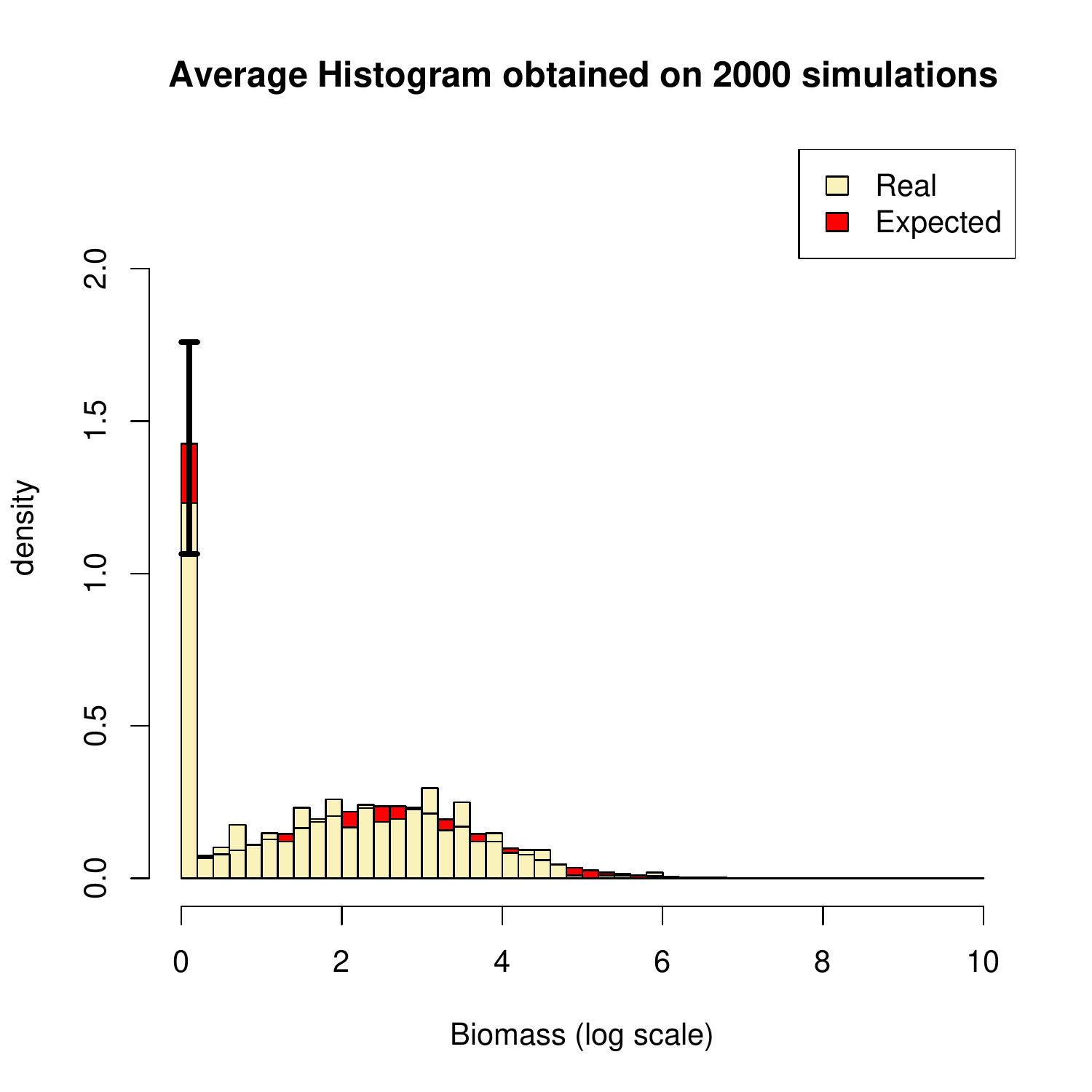}
    \caption{Comparisons between Sunflower Starfishes dataset and averaged histogram (1000 simulated datasets at $\hat{\theta}^{Sun}$)}
    \label{fig:sunstarfit}
  \end{minipage}
\end{figure}

 A visual diagnosis of the goodness of fit is very
informative. According to the RLOL model, data are drawn from a
mixture and we cannot add directly a density line on the histograms
of figures \ref{Fig:Urchinhist} and \ref{Fig:SunStarhist} since the
zero ordinate of these figures is somewhat artificial : it depends
on the width of the histogram bins and has been chosen so that the
overall cumulative greyed surface is 100\%. The expected histograms
presented in figures \ref{fig:urchinfit} and \ref{fig:sunstarfit}
have been obtained using 1000 replications of the model with the
same design at $\hat{\theta}$, and averaging the 1000 generated
histograms. Obviously the obtained model histogram (averaging all
the random effects) is smoother than the empirical distribution. The
observed number of zeros falls below the expected number but within
the 90\% confidence interval for each species (as indicated by the
vertical line on figures \ref{fig:urchinfit} and
\ref{fig:sunstarfit}) and the overall shape of the distribution fits
quite well the data in both cases.

\medskip

\subsubsection{Confidence intervals}

 Relying on proposition \ref{prop:Fisher}, the asymptotic covariance matrices are evaluated at those maximum likelihood arguments~: {\small
\begin{eqnarray*}
\left(
\begin{array}{c}
var(\hat{a}^{Urch},\underline{\mathbf{Y}}) \\
var(\hat{b}^{Urch},\underline{\mathbf{Y}}) \\
var(\hat{c}^{Urch},\underline{\mathbf{Y}}) \\
var(\hat{d}^{Urch},\underline{\mathbf{Y}})%
\end{array}%
\right)  &=&\left(
\begin{array}{c}
0.0587 \\
0.1020 \\
1.6804 \\
14.4793%
\end{array}%
\right)  \\
Corr(\hat{\theta}^{Urch},\underline{\mathbf{Y}}) &=&\left( {\small
\begin{array}{cccc}
1 & 0.825 & 0.035 & 0.058 \\
0.825 & 1 & 0.036 & 0.081 \\
0.035 & 0.036 & 1 & 0.936 \\
0.058 & 0.081 & 0.936 & 1%
\end{array}%
}\right)
\end{eqnarray*}%
} and {\small
\begin{eqnarray*}
\left(
\begin{array}{c}
var(\hat{a}^{Sun},\underline{\mathbf{Y}}) \\
var(\hat{b}^{Sun},\underline{\mathbf{Y}}) \\
var(\hat{c}^{Sun},\underline{\mathbf{Y}}) \\
var(\hat{d}^{Sun},\underline{\mathbf{Y}})%
\end{array}%
\right)  &=&\left(
\begin{array}{c}
0.2555 \\
0.3003 \\
0.2609 \\
0.0894%
\end{array}%
\right)  \\
Corr(\hat{\theta}^{Sun},\underline{\mathbf{Y}}) &=&\left(
\begin{array}{cccc}
1 & 0.902 & -0.055 & -0.046 \\
0.902 & 1 & -0.056 & -0.023 \\
-0.055 & -0.056 & 1 & 0.906 \\
-0.046 & -0.023 & 0.906 & 1%
\end{array}%
\right)
\end{eqnarray*}%
}
 Essentially only $\hat{a}$ and $\hat{b}$
(resp $\hat{c}$ and $\hat{d}$) are correlated.

\begin{figure}[h!]
  \begin{minipage}[b]{0.45\linewidth}
    \centering \includegraphics[scale=0.3]{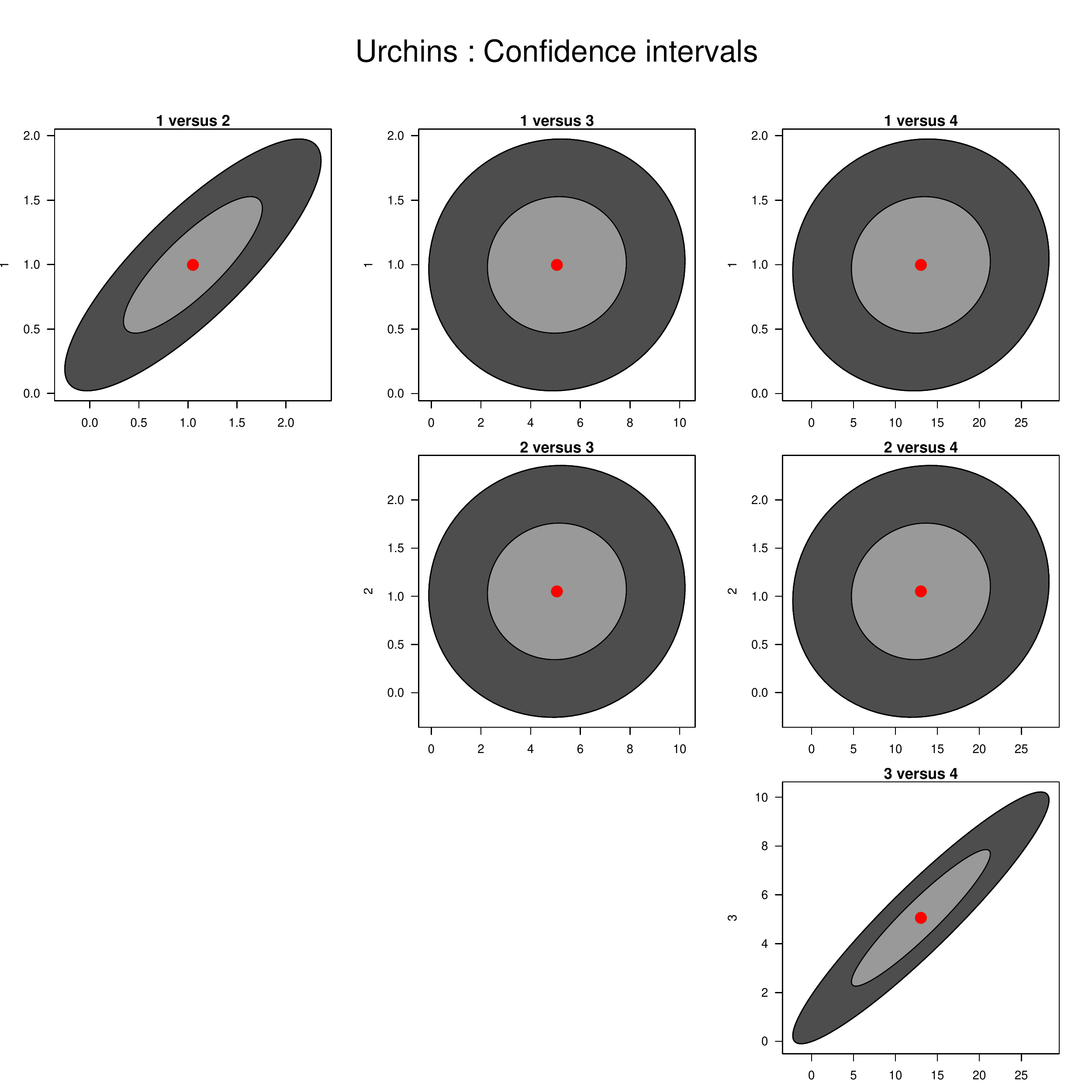}
    \caption{The ligthest ellipse corresponds to  90\% confidence ellipsoid and the darkest one is 99.96\% and contains 90\% of the simulated values.}
    \label{fig:ellipseurchins}
  \end{minipage}\hfill
  \begin{minipage}[b]{0.45\linewidth}
    \centering \includegraphics[scale=0.3]{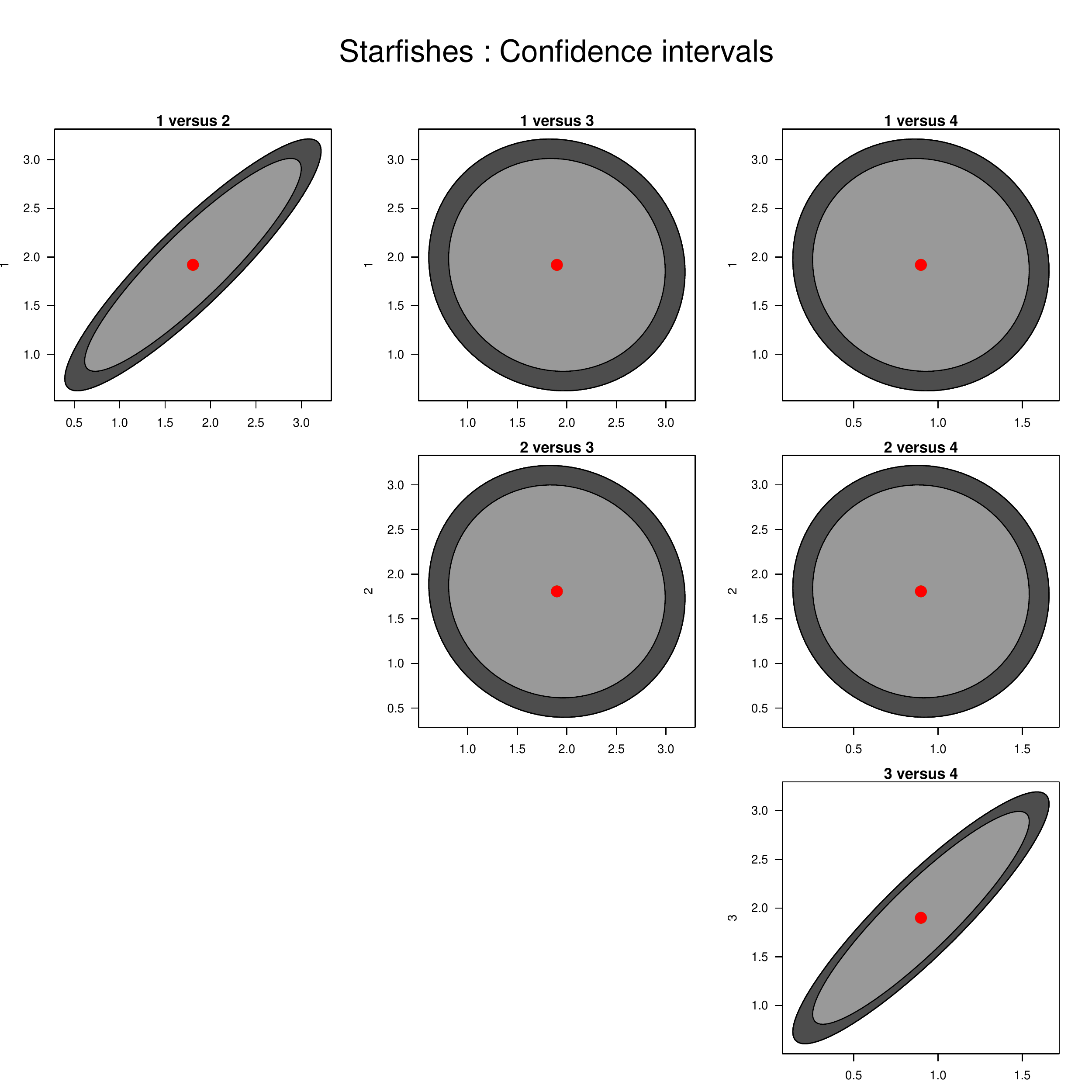}
    \caption{The ligthest ellipse corresponds to  90\% confidence ellipsoid and the darkest one is 94\% and contains 90\% of the simulated values.}
    \label{fig:ellipsesunstars}
  \end{minipage}
\end{figure}

To evaluate the actual coverage of confidence regions in the present
sampling conditions (that may be far from asymptotics), 16000
simulations were launched, assuming the same number of strata and
the same number of data points per stratum as the urchin catches
(resp. sunflower starfishes) with $\hat{\theta}$ as hypothetic true
parameter, thus disregarding possible bias. As a practical working
conclusions, Figures~\ref{fig:ellipseurchins} and
\ref{fig:ellipsesunstars} show how to correct theoretical
asymptotical confidence intervals. The results are quite different
from one dataset to the other.
\begin{itemize}
\item On Urchins dataset, to get an actual 90\% confidence region, we must expand as far as the asymptotic ellipse corresponding to a 99.964\% normal approximation
as shown in Figure~\ref{fig:ellipseurchins}.
\item On Sunflower Starfish dataset, things work better and the 94\%
asymptotical confidence interval is quite a good surrogate for an
actual 90\% confidence region!
\end{itemize}

To understand Table~\ref{tab:ConfInt}, we suggest to consider the
median column as the reference confidence interval (based on
simulation/ EM re-estimation). The right column gives bootstrap+ EM
re-estimation. We notice that the Bootstrap approach is completely
unappropriate for our model. The estimation is clearly biased with a
shift to the right (verified on simulations not shown here) although
we tried to correct bias as proposed in \cite{Hesterberg04}. The
width of confidence intervals are underestimated for both species
and does not even contain the $\hat{\theta}$-value. The hierarchical
structure of the model may explain part of this bad behavior of
bootstrap method but this would need further investigations not in
the scope of this paper. The left column of Table~\ref{tab:ConfInt}
exhibits two different behaviors according to the species
considered.\begin{itemize}
                                                          \item
The asymptotic variance of maximum likelihood parameters
under-estimate strongly the true sampling characteristics in the
Urchin case. This may be due to the large numbers of zero's for that
species: consequently relatively less non zero data remain for the
$\rho'$s (inverse of patch abundance) and the estimation of $c$ and
$d$ that rule the between units variation of $\rho$'s may become
difficult.
                                                          \item
The Sunflower Starfishes case exhibits much better properties
regarding the approximation of the covariance matrix. For this
species, less zeros data occur and we guess that enough information
is made available in the sample to get correct estimations.
                                                        \end{itemize}

\begin{table}
  \centering
  \begin{tabular}
[c]{|c|c|c|}\hline
\multicolumn{3}{|c|}{ $90\%$ Confidence  Intervals}\\
(asymptotic) & (via simulation) & (via booststrap)\\ \hline \hline
\multicolumn{3}{|c|}{Urchins case}\\ \hline
$0.587<a<1.384$ & $0.335<a<1.637$ & $0.72<a<1.00$\\
$0.496<b<1.547$ & $0.163<b<1.880$ & $0.61<b<1.03$\\
$2.827<c<7.092$ & $1.476<c<8.443$ & $1.23<c<4.37$\\
$6.387<d<18.905$ & $2.419<d<22.872$ & $1.68<d<10.89$\\ \hline \hline
\multicolumn{3}{|c|}{Starfishes case}\\ \hline
$ 1.087<a< 2.750  $ & $1.294 <a< 2.951$ & $1.217<a<1.859$\\
$ 0.905 <b<  2.708$ & $ 1.198<b<3.141$ & $0.938<b<1.663$\\
$  1.059<c<  2.740$ & $1.344<c<3.182$ & $1.147<c<2.035$\\
$ 0.406<d<  1.390$ & $0.558<d<1.559$ & $0.347<d<0.858$\\ \hline
\end{tabular}
  \caption{Comparison of the asymptotic 90\% confidence interval
  with the one obtained by simulation for each parameter component for both species}
  \label{tab:ConfInt}
\end{table}

\begin{figure}[h!]
  \begin{minipage}[b]{0.45\linewidth}
    \centering \includegraphics[width=8cm]{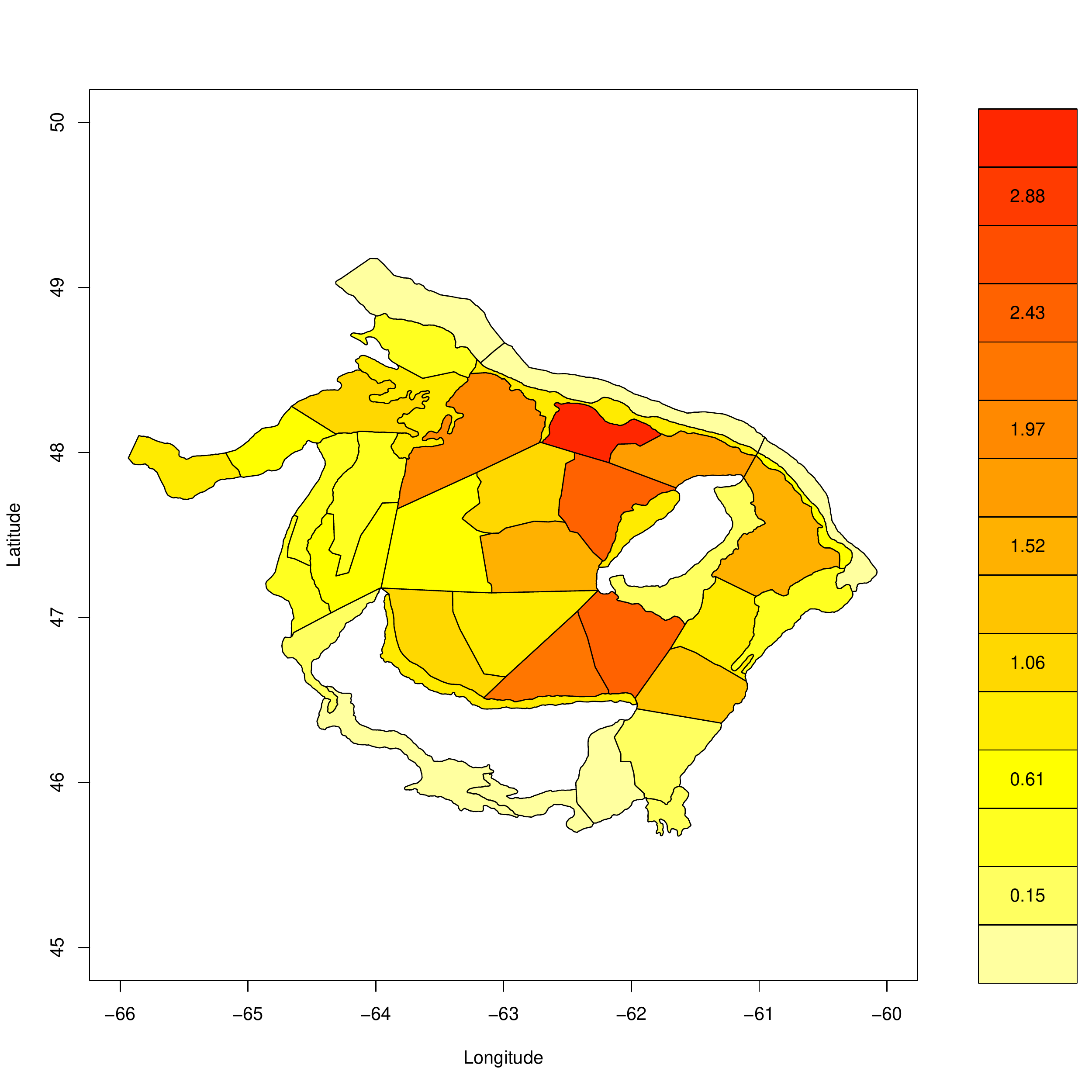}
    \caption{Predictions of the random effects $\mu_s$ in each stratum  correspond to the expected number of clumps collected during a measurement with standardcatching effort.}
    \label{Fig:predmu}%
  \end{minipage}\hfill
  \begin{minipage}[b]{0.45\linewidth}
    \centering \includegraphics[width=8cm]{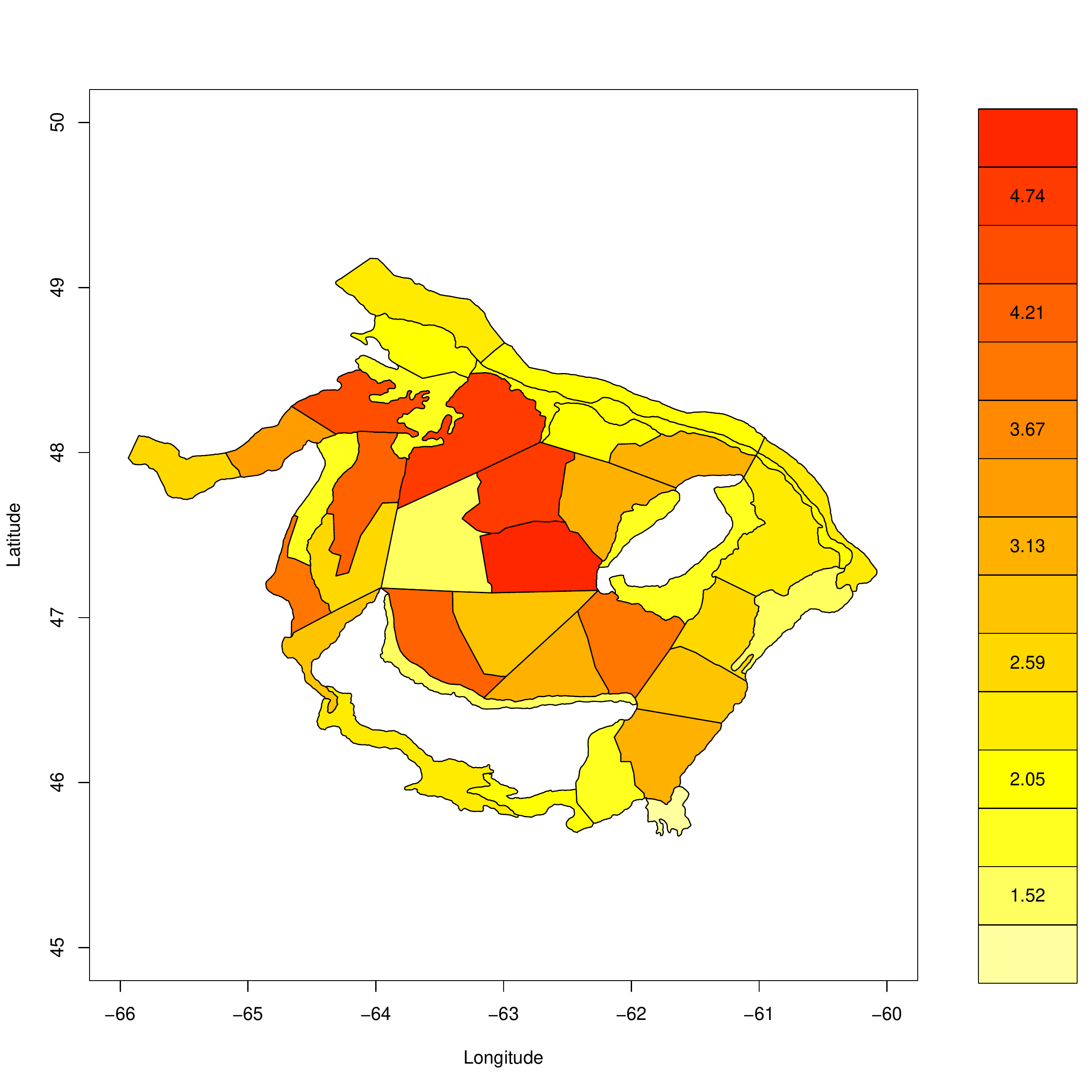}
    \caption{Predictions of the inverse of $\rho_s$  in each stratum.
      These quantities give the expected biomass to be collected within a
      clump.}
    \label{Fig:predrho}%
  \end{minipage}
\end{figure}

Figures \ref{Fig:predmu} and \ref{Fig:predrho} present the
predictions for the random effects in each stratum.

\subsubsection{Validation of the gamma assumption for random effects}
\label{sec:gammaassumption} 
\begin{figure}[h!]
  \centering \includegraphics[scale=0.4]{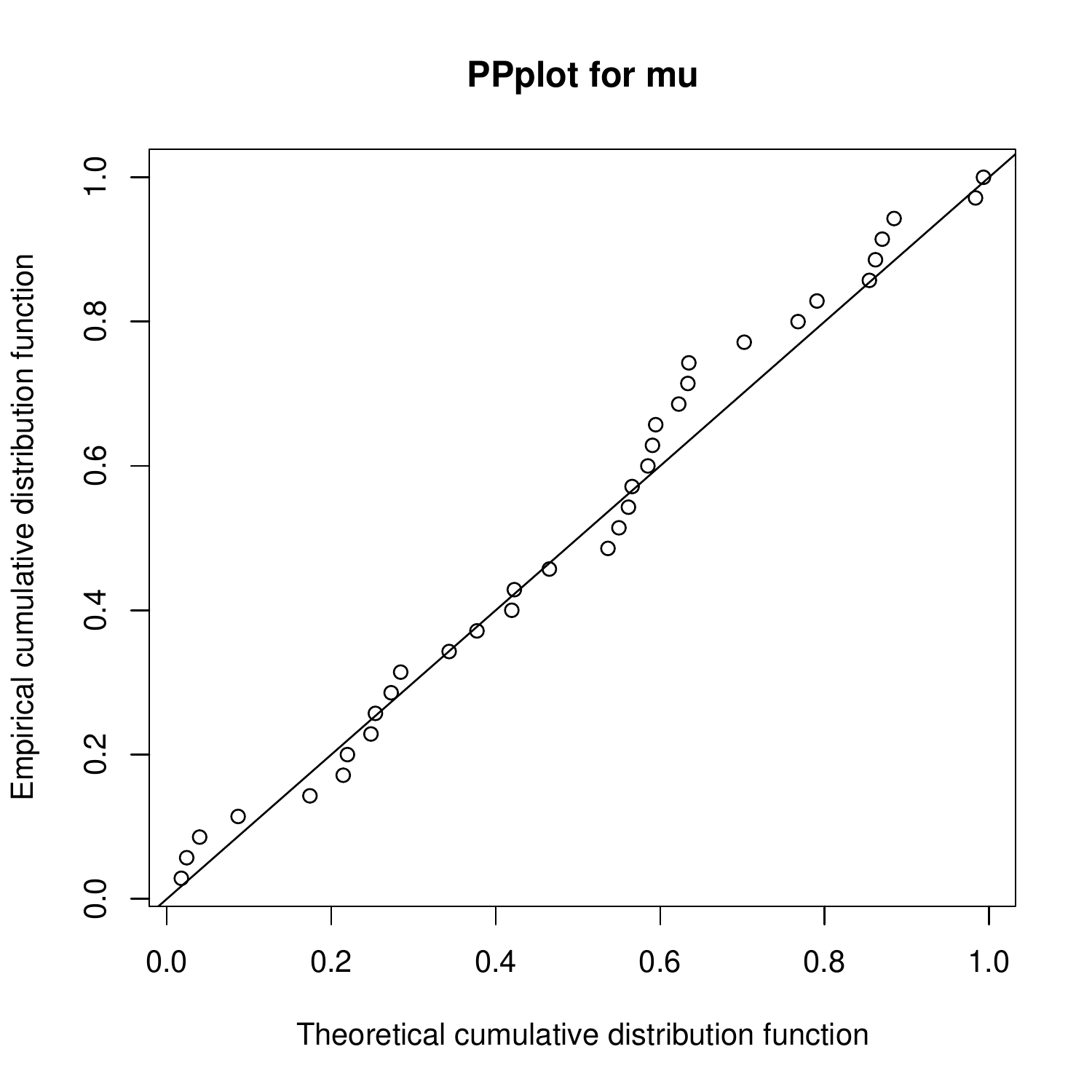}
  \caption{pp-Plot with estimates of $\mu_s$ versus a fitted gamma distribution.}
  \label{Fig:ppplotmu}%
\end{figure}

\begin{figure}[h!]
  \begin{minipage}[b]{0.45\linewidth}
    \centering \includegraphics[scale=0.3]{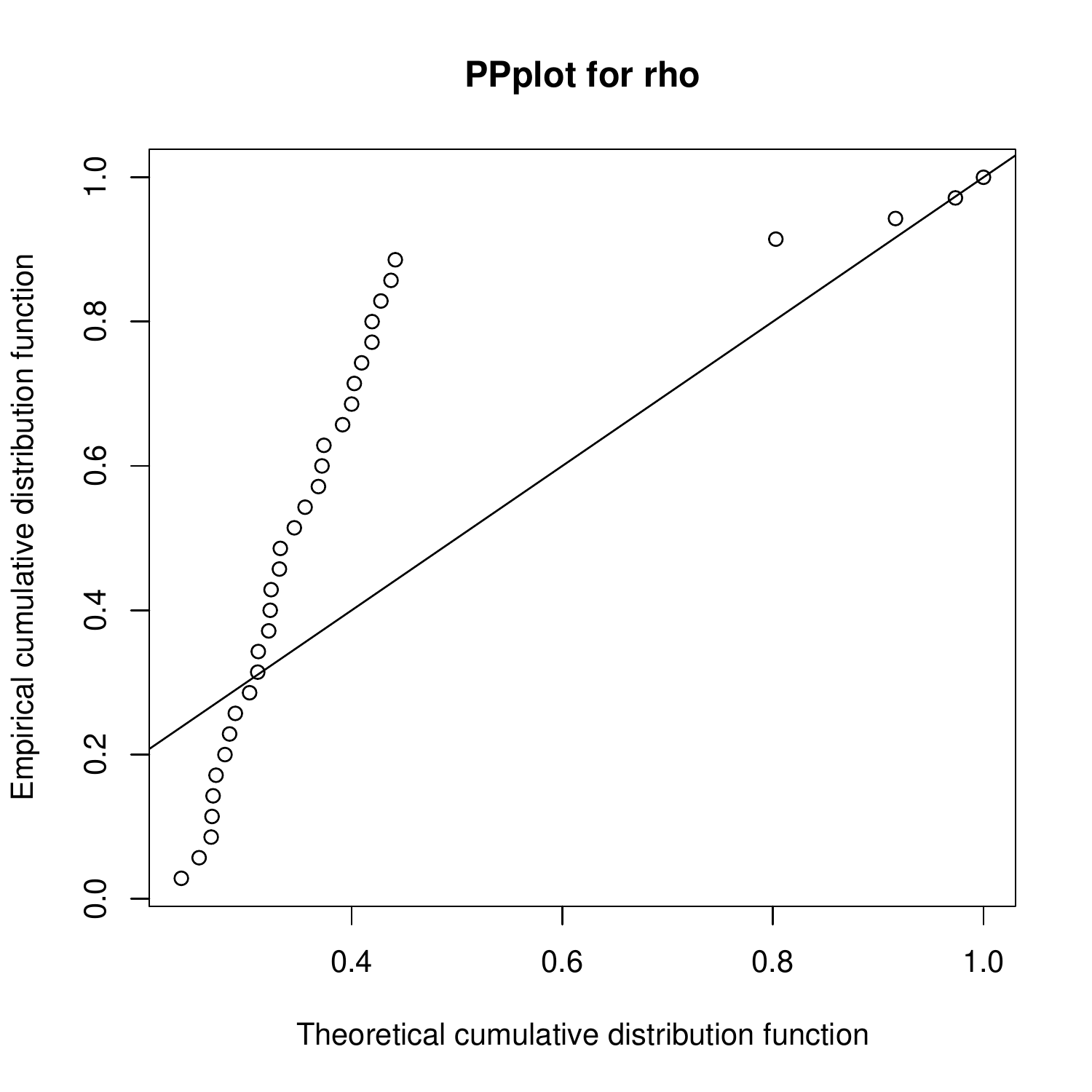}
    \caption{pp-Plot with estimates of $\rho_s$ versus a fitted gamma
      distribution. The extremal points correspond to strat with at least $75\%$ of zeros}
    \label{Fig:ppplotrho}%
  \end{minipage}\hfill
  \begin{minipage}[b]{0.45\linewidth}
    \begin{center}
      \includegraphics[scale=0.3]{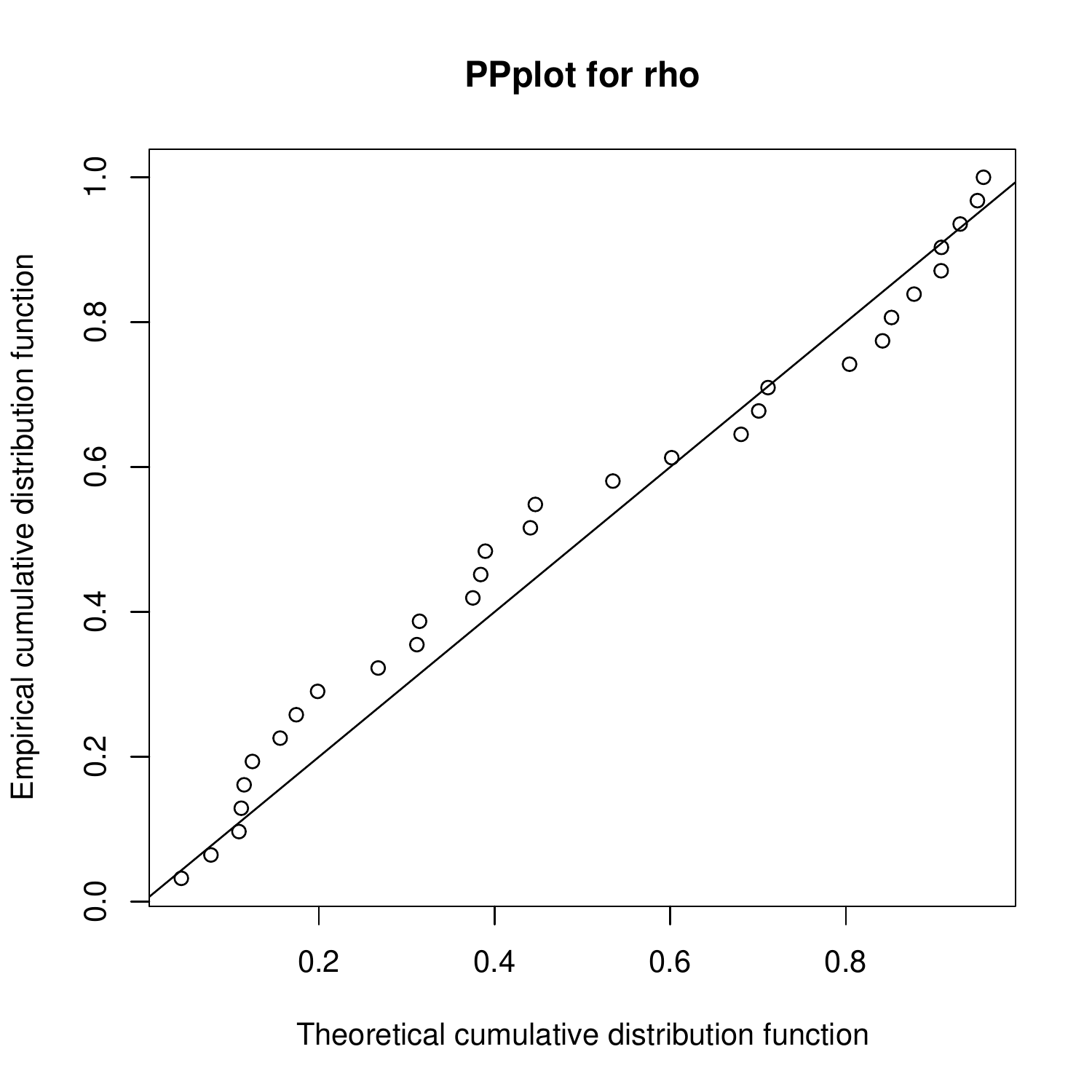}
    \end{center}
    \caption{pp-Plot with estimates of $\rho_s$ against a fitted gamma
      distribution after excluding the four  outliers.}
    \label{Fig:ppplotrho2}%
  \end{minipage}
\end{figure}

We have assumed that the random effects
$\mu$ and $\rho$ were distributed according to gamma distributions.
This choice was essentially made for technical convenience because
conjugate properties make the estimation easier. The validity of
this assumption can be checked by considering random effects as
fixed and estimate them independently in each stratum. Figures
\ref{Fig:ppplotmu} and \ref{Fig:ppplotrho} present a pp-plot of
empirical versus estimated probability distributions for $\mu$ and
$\rho$.

The pp-plot for $\mu$ suggests that the gamma distribution is
appropriate (Fig~\ref{Fig:ppplotmu}); this is not true of the gamma
pp-plot for $\rho$ (Fig~\ref{Fig:ppplotrho}). First there are only
36 points estimates because 2 strata are empty and $\rho$'s for
these strata are not defined. Second the probability plot does not
adjust to a straight $45$ degrees line. Looking more closely at four
extreme points in the $\rho$ pp-plot, we found that they come from
strata with less than two non-zero data points. Excluding these 4
points produces the much more acceptable fit of
Figure~\ref{Fig:ppplotrho2}.

\subsection{Simulations Studies}
The previous section showed different behaviors depending on the
species : the EM procedure provides rather reliable estimates for
the starfish RLOL statistical features but not for the Urchin ones.
The purpose of this section is to check the role of the sampling
designs. Simulation studies are performed to explore the quality of
the EM estimation procedure and to check the actual coverage of the
asymptotic variance-covariance matrix approximation.
\subsubsection{Simulation design}

For a given set of parameters $\theta=(a,b,c,d)$, we draw 1000
samples according to RLOL  model given in eq \ref{eq:RLOL1} with a
number $S$ of strata and $M$ measured points per each stratum. $S$
has been chosen varying as $k^{2}$ with $k=3,4,5,6,8,10,12,15$ and
$M=5,10,15,20,25,30,40$.

For each simulation, the estimation procedure depicted in section
\ref{sec:EM} yields one point estimate and one estimation of the
asymptotic covariance matrix. Assuming that the asymptotic
approximation holds and using a normal approximation, confidence
intervals can be given for the \emph{true} value. As we work within
a simulation context, the \emph{true} value is known and one can
compute the actual proportion of samples for which the asymptotic
confidence interval covers the \emph{true} value.

\subsubsection{RLOL Results}
The simulation study is achieved for two values of parameters
$\theta$ corresponding to the two applications developped in section
\ref{sec:realdata}. We choose $\theta^{Urchin}=(1,1,5,13)$  and
$\theta^{Sunstars}=(1.9,1.8,1.9,0.9)$  as \emph{true} parameter
references for the simulations. We first present a study of the bias
and then an investigation of the actual coverage of confidence
intervals.

\begin{figure}[h!]
  \centering \includegraphics[scale=0.3]{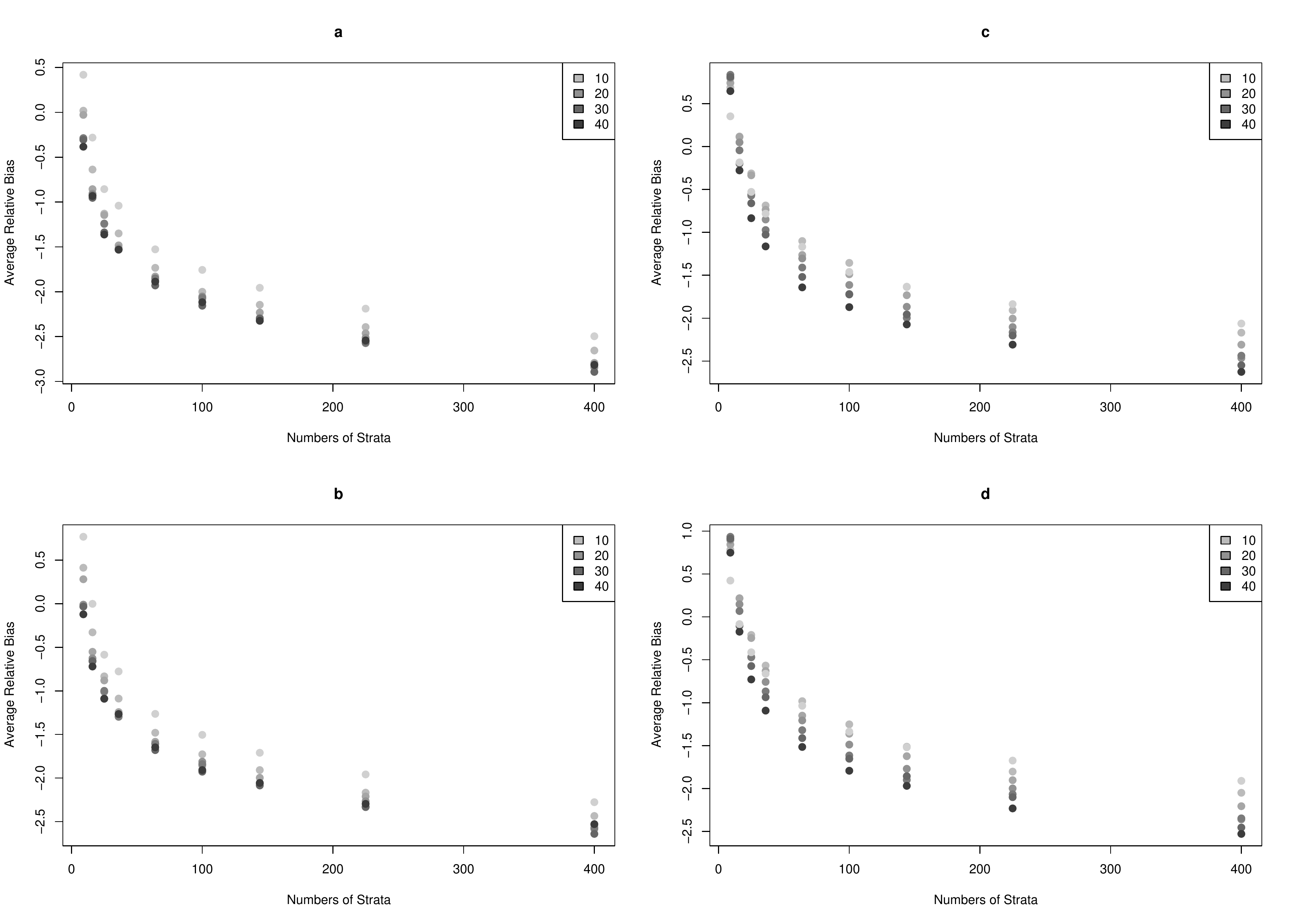}
  \caption{Urchins : Average relative bias in log scale depending on the number
of strata and the number of measure points.}
  \label{fig:Bias_Urchin}
\end{figure}

\begin{figure}[h!]
  \centering \includegraphics[scale=0.3]{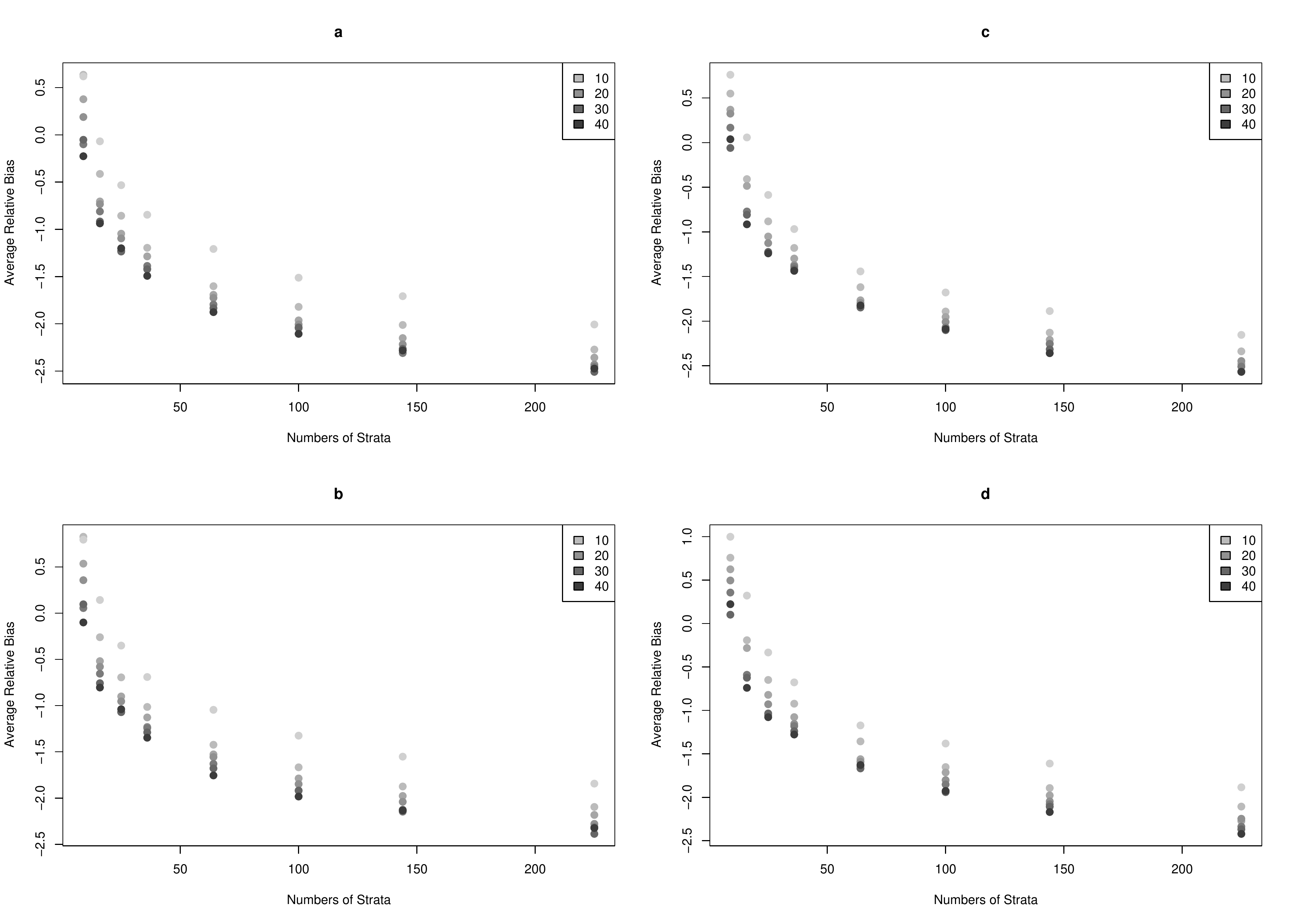}
  \caption{Starfish : Average relative bias in log scale depending on the number of strata and the number os measure points.}
  \label{fig:Bias_Stunstar}
\end{figure}

{\bf Bias study}\\
We can study the bias by simulation according to the numbers of
strata and the number of measure points within strata. Figures
\ref{fig:Bias_Urchin} and \ref{fig:Bias_Stunstar} present the
results for relative bias obtained with 1000 simulations in each
configuration. As expected it decreases quickly with the number of
strata and only marginal amelioration is obtained as soon as the
number of data per stratum becomes reasonable.

\begin{figure}[h!]
 \begin{center}
   \begin{minipage}[b]{0.45\linewidth}
     \begin{center}
       \centering \includegraphics[scale=0.45]{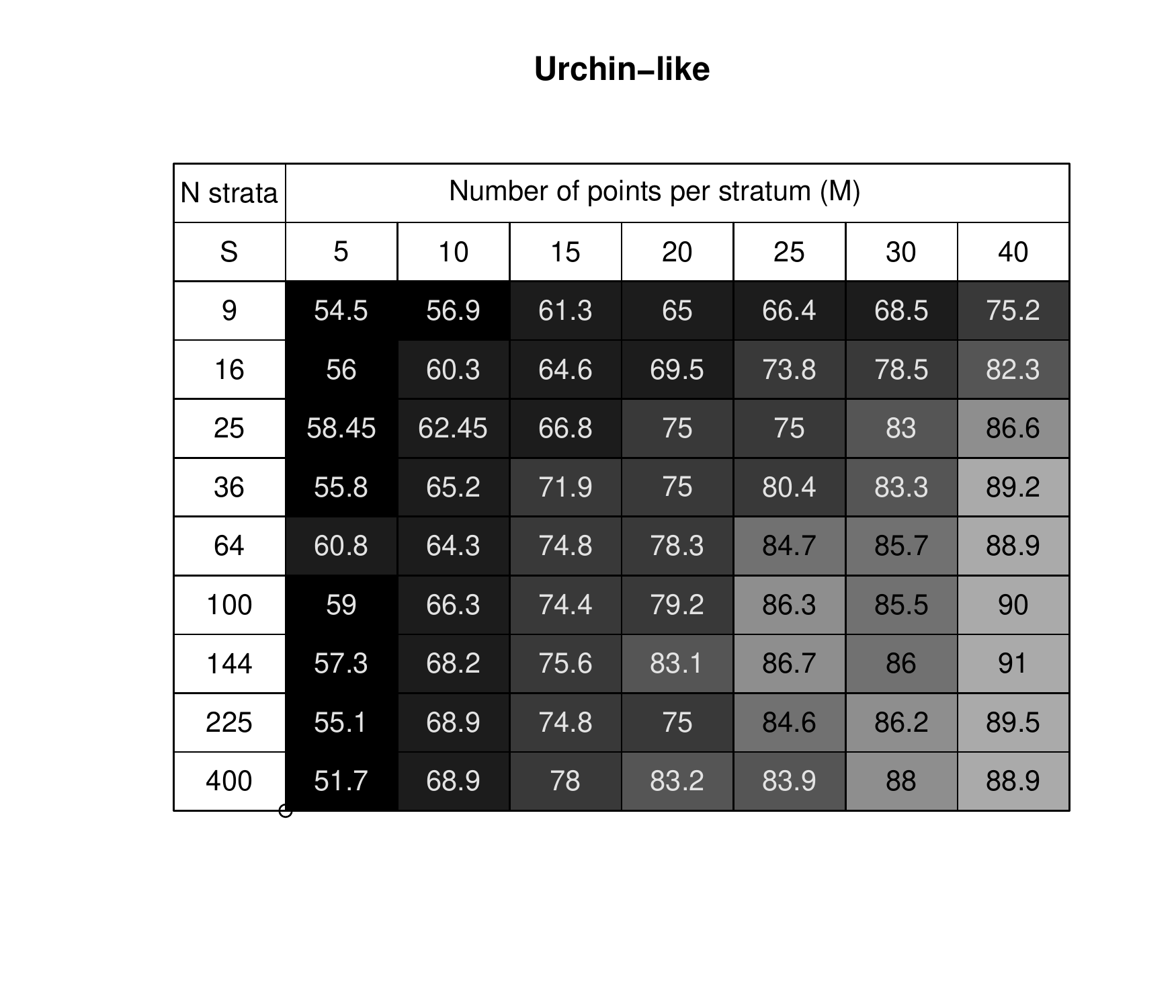}
       \caption{Urchin-like case. Effective proportion of 90\% confidence intervals that cover the true value. Shading in particular cells reflects the degree of overlap: M-S combination that produces confidence intervals that are too liberal are in black whereas the lightest grey shade reflects confidence intervals that properly characterize parameter uncertainty}
       \label{Fig:ICUrchin}%
     \end{center}
   \end{minipage}\hfill
   \begin{minipage}[b]{0.45\linewidth}
     \begin{center}
       \centering \includegraphics[scale=0.45]{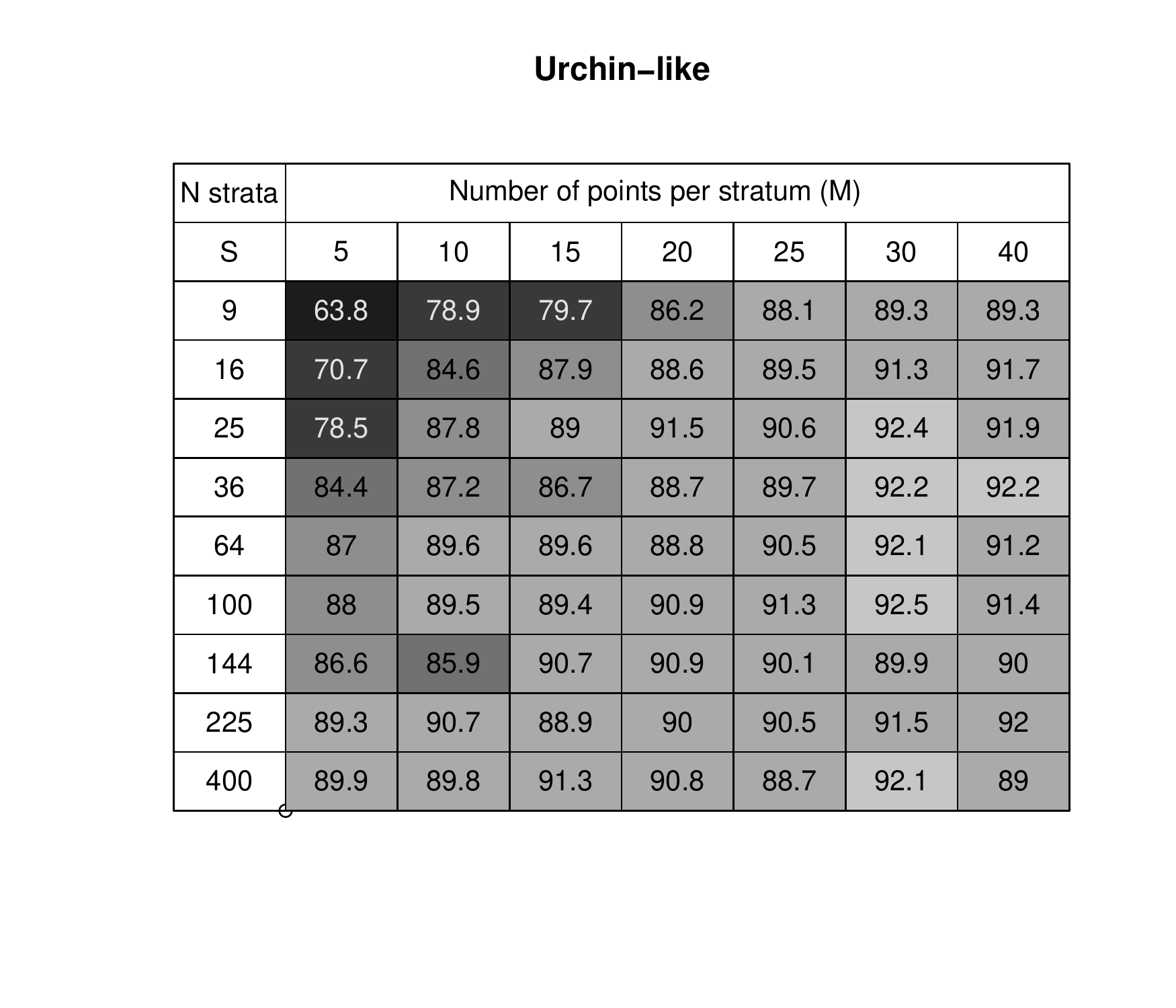}
       \caption{Sunstar-like case. Effective proportion of 90\% confidence intervals that cover the true value. Shading in particular cells reflects the degree of overlap: M-S combination that produces confidence intervals that are too liberal are in black whereas the lightest grey shade reflects confidence intervals that properly characterize parameter uncertainty.}
       \label{Fig:ICSunstar}%
     \end{center}
   \end{minipage}
 \end{center}
\end{figure}

\medskip
{\bf Confidence intervals study}\\ Using 1000 simulations in each
cell, the empirical proportion of the asymptotic $90\%$ confidence
ellipsoids that cover the \emph{true} value is given in
Figures~\ref{Fig:ICUrchin} and \ref{Fig:ICSunstar}. With 1000 trials
in a binomial distribution with probability $p$ of success, a
confidence interval for $p=0.90$ is approximatively $[88\%, 92\%]$ :
cells from Figures~\ref{Fig:ICUrchin} and \ref{Fig:ICSunstar} that
belongs to that interval have been colored in light grey. Results
about confidence intervals strongly depend on the value of $\theta$.
The asymptotic approximation seems quite satisfying for
$\theta^{Sunstars}$ : the asymptotical conditions are quickly
fulfilled and the design of the case study seems acceptable.  For
$\theta^{Urchin}$ however, the present design should be strongly
re-enforced (up to 40 points per stratum with 36 strata!) before
yielding acceptable estimations, and confidence regions based on
asymptotical theory are definitely too optimistic.

These two sets of parameter recover two very different situations :
the larger number of zeros in the Urchin case may render the
estimation procedure more difficult than in the Starfish situation.
However one should note that the difference is not markedly
pronounced : $34 \%$ instead of $24\%$! Such a simulation study
shows that the quality of variance covariance matrix estimation used
to build an ellipsoid of confidence behaves has to be checked
through this simulation approach by instance to verify whether  the
asymptotic conditions are  fulfilled and that the analyst should
beware of overconfidence.

\section{Conclusion and Perspectives}

The following conclusions have been reached:

\begin{enumerate}
\item  Compound Poisson distributions can conveniently represent the presence
of a large number of zeros and a skewed distribution of non-zero values. To
deal the occurrence of zero-inflated data, very parsimonious models can be
designed (with two parameters only)~: a Poisson random sum of independent
geometric random variables in the discrete case and with exponential random
variables in the continuous one. They offer an alternative to the traditionnal
delta gamma models and behave coherently when changing the scale of the catch
effort, thanks to the Poisson process underpinning the model.

\item  Compound Poisson distributions can be interpreted using a
hierarchical framework. They describe the data collection involved
in sampling individuals gathered in (latent) patches drawn from the
homogeneous Poisson process with \emph{abundance} tuned by the
distributional parameter of the random components of the Poisson
sum. The introduction of a random effect structure at the top of the
hierarchy is straightforward and accommodates non homogeneity among
strata that are themselves considered as homogeneous units. Such
designs with random effects and data with extra zeros are commonly
encountered in ecological analyzes, but gamma random effects are yet
rarely advocated : variation between strata is typically modeled
using a normal (or lognormal) distribution because its sufficient
statistics match the commonsense interpretation of mean and
variance. However, gamma random effects allow for partial conjugate
properties with the compound Poisson model for zero-inflated data.
Beyond this theoretical convenience, the parameters of the gamma
distribution are well estimated in the Starfish like simulation
examples and they can describe the entire range of variability
between units for the real case study.

\item  Independence between the latent features $\rho$ and $\mu$ has been a priori
assumed for the random effects between units. This absence of prior
correlation is quite a stringent hypothesis as we might expect
$\rho$ and $\mu$ to covary~(e.g, low non-zero realized abundance
could stem from either a small $\mu$ or a large $\rho$). Working
with a gaussian copula for a joint bivariate distribution for the
couple ($\mu,\rho)$ is a bad remedy,\ because we would have lost the
conjugate properties and increased computational load. To keep
partial conjugacy , a better idea is considering the natural
extension of the gamma family, but such bivariate distributions are
rather restrictive since they can only take into account positive
correlation and need that the two marginals share the same shape
parameter. However such a model would remain parsimonious with 4
parameters: one is gained to depict correlation and one is lost to
depict the marginals'shape. The issue of correlation has been
addressed in \cite{Anceletthese2008} who proved via simulation that
the correlation between $\rho$ and $\mu$ has little bearing on the
property we are ultimately trying to predict in practice, i.e. the
realized biomass in a tow. Finally, the correlation indicates that
the latent variables $\rho$ and $\mu$ are model concepts that should
themselves not be overinterpreted; they don't actually characterize
the true size and number of organism patches.

\item  Stochastic EM inferential techniques (with importance sampling for the
non explicit expectation steps) require a modest computational
effort since the random effects are taken partially conjugate with
the compound Poisson distributions. Auxiliary importance
distributions can be proposed by careful inspection the structure of
the joint distribution of the latent variables and integrating out
as much as can\ analytically be done. Much advantage is taken from
conditional independence, especially when computing the Fisher
information matrix by re-sampling with the simulated missing data
that have been previously generated to evaluate the maximum
likelihood estimate. However, the value of results given here
depends on the errors involved with the use of maximum likelihood
asymptotic formula on one hand and on the precision of Monte Carlo
sampling algorithms on the other hand. Due to the multidimensional
nature of the latent variables to be simulated , the variability
between several trials of the importance sampling techniques when
evaluating the information matrix (and its inverse) can be important
enough, especially when few data makes a rather flat likelihood
function.

\item   Asymptotic errors bounds need to be checked and corrected if necessary.
We relied on a simulation study to get a more reliable idea of their
ranges. The simulated sets of zero-inflated data show that, in the
Starfish case, one can readily trust the confidence intervals based
on the information matrix while in the Urchin case, one should
beware of being overconfident. The asymptotic conditions may not be
encountered rapidly. For the Starfish case study, the design allowed
a reasonable estimation of the RLOL model features. For the other
species with a 10\% higher probability of getting zero values,
safisfying precision estimates with 40 strata need at least
collecting 40 data points per stratum before the confidence coverage
gets reasonably close to its theoretically recommended approximate
value. Because 1600 stations represents generally unrealistically
large sampling effort for a marine bottom-trawl survey in that
Urchin example, statisticians need to inform practitioners (before
launching the data collection) about possible underestimation of
uncertainty.

\item  Covariates for the fixed effect of environmental variable
(depth, temperature and habitat type) could be added to the model,
potentially enhancing ecological interpretation of the observed
patterns in organism abundance and distribution. However, it may
bring a lot of additional burden during the inferential computations
since many of the conjugate properties would be lost. For the same
reasons, non exchangeable strata (with for instance an intrinsic CAR
structure on the top of the hierarchy as described in
\citet{Ban2004}) have not been considered here. Simple (low
dimensional) importance sampling should be replaced with brute force
Hastings Metropolis techniques \citep{Hastings70}. In such a
context, it may be worthwhile to work on encoding prior knowledge
\citep{Kadane+98} into probability distributions and switch the
problem into a Bayesian framework \citep{Berger85}, relying on
ready-made tools such as WinBugs for inference \citep{Spie+2000}.

\item  In the case study, the random effect models with compound Poisson
distribution for the occurrence of zero-inflated data fit the data
well and allow transfer of information between strata to help
predict in data-poor units. Its hierarchical structure favors
discussion between ecologists and statisticians, and helps query its
interpretation in term of ecological situations with extra zeros.
\end{enumerate}

\bibliographystyle{plain}
\bibliography{biblio}

\appendix
\begin{center}
    APPENDICES
\end{center}

\section{Compound
Poisson process characteristic function} \label{an:fc}
 When $X$ is real valued, we denote by $\hat{f}$ the
Fourrier transform\footnote{For non negative integer valued random
variables $X$\ the
probability generating function $P(z)=%
{\displaystyle\sum\limits_{0}^{\infty}}
\Pr(X=n)z^{n}$ is the corresponding machinery for handling discrete
distributions~: the same results can be found in this case by setting the
change of variables $z=e^{i\omega}$} of $f$ (i.e the characteristic function
of $X$)~:
\[
\hat{f}(\omega)=E(e^{i\omega X})
\]

From equation~\ref{compounddef}, the compound Poisson distribution $g$ is such
that~:%
\begin{equation}
\hat{g}(\omega)=%
{\displaystyle\sum\limits_{n=0}^{\infty}}
e^{-\mu}\frac{\mu^{n}}{n!}\left(  \hat{f}(\omega)\right)  ^{n}=e^{-\mu
(1-\hat{f}(\omega))} \label{compoundfc}%
\end{equation}

This equation exhibits the infinite divisibility property of $Y$
with regards to parameter $\mu$, which offers a nice conceptual
interpretation when returning to the marked Poisson process
underneath this stochastic construction~: the resulting quantity $Y$
is obtained by collecting a random number of primarily (hidden)
batches $X_{i}$ distributed at random with intensity $\mu$. Such a
conceptual latent process of aggregates would be intuitive for many
ecologists. Conversely, one can easily check by writing the
logarithm of their characteristic functions, that traditional models
for zero-inflated data (think for instance of the delta-gamma model
or the Zero-Inflated Poisson model such as \cite{Ridout+98}) lack of
coherence for adapting to a change of the scale in the experiment.

Among the many choices for the probability distribution $f$\ of the
random mark of the sum, this paper focuses, for parsimony and
realism, on the exponential distribution for $X$ (continuous case)
that is~:
\[
f(x)=\rho e^{-\rho x}%
\]
$\ $ so that $\hat{f}(\omega)=\frac{\rho}{\rho+i\omega}$ and $\hat{g}%
(\omega)=e^{-\mu(\frac{i\omega}{\rho+i\omega})}$ . For the discrete case, we
suggest the \ corresponding geometric distribution~: $f(x)=1_{x>0}%
\times(1-r)\times r^{x}e^{iwx}$ leading to $\hat{f}(\omega)=\frac
{1-r}{1-re^{i\omega}}$ and $\hat{g}(\omega)=e^{-\mu\left(  \frac
{r(1-e^{i\omega})}{1-re^{i\omega}}\right)  }$for the exponential compound
Poisson count model.

\section{Initialization of the Newton-Raphson algorithm}

\label{an:NR} 
\begin{figure}[htb]
\centering{\scalebox{0.35}{\includegraphics*{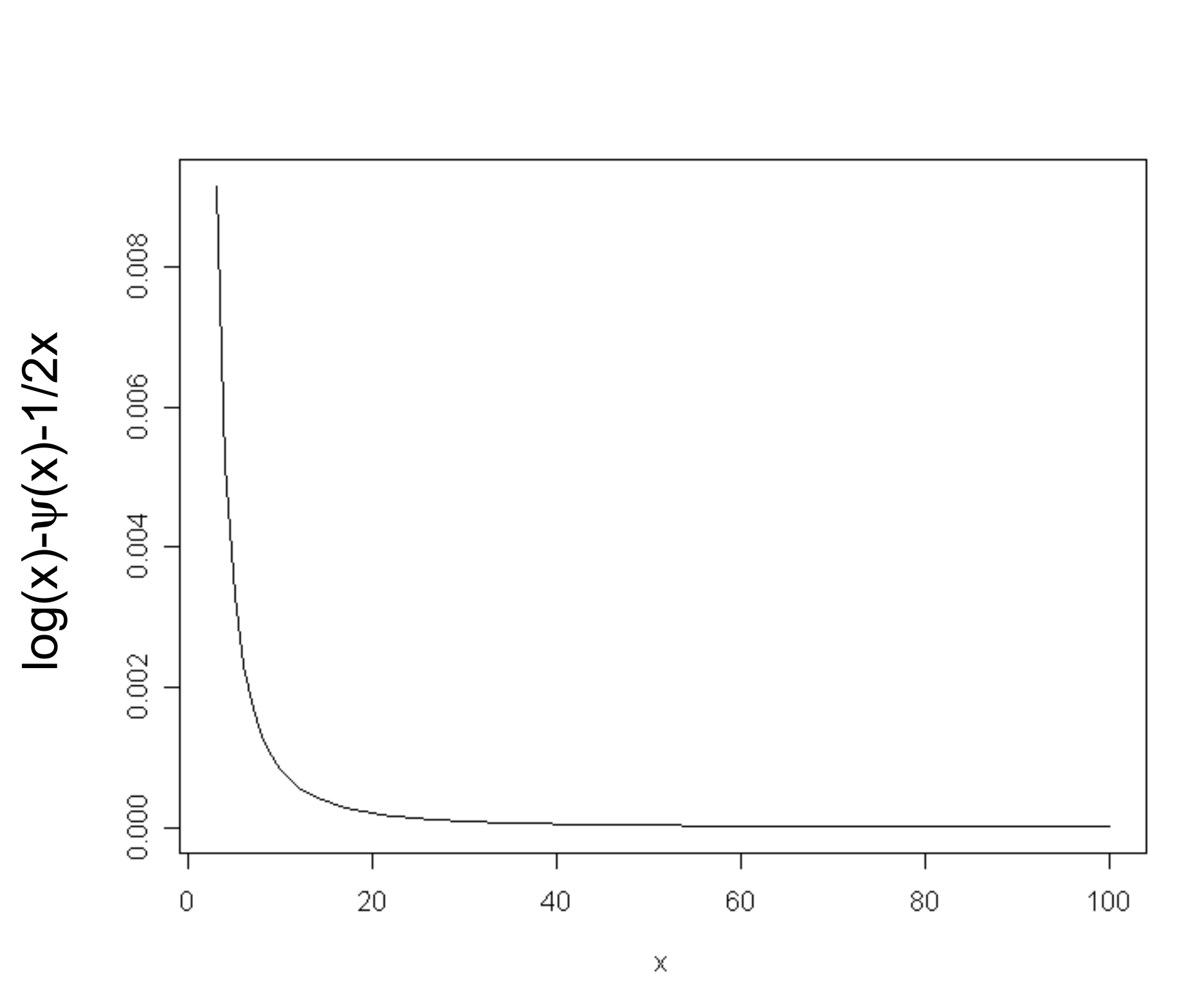}}}\caption{
Difference between $log(x)-\psi(x)$ and $1/2x$}%
\label{Fig:DifflogPsi}%
\end{figure}

The main point on Newton-Raphson algorithm consists in
choosing a good initial point. In this paper we use this algorithm
to find the zero of
\[
\ln(a)-\psi(a)-C=0
\]
Note that function $\psi$ verifies the following asymptotic
series'~expansion \citep{Abramowitz1964}~:
\begin{align*}
\psi(x)  &
\underset{x\rightarrow\infty}{\sim}\ln(x)-\frac{1}{2x}-\sum
_{n=1}^{\infty}\frac{B_{2n}}{2n\,x^{2n}}\\
&  \underset{x\rightarrow\infty}{\sim}\ln(x)-\frac{1}{2x}-\frac{1}{12\,x^{2}%
}+\frac{1}{120\,x^{4}}+\ldots
\end{align*}
The convergence is very fast (see Figure~\ref{Fig:DifflogPsi}) so
that we choose to initiate Newton-Raphson algorithm with
$x_{0}=\frac{1}{2C}$.

\section{Computation of the moments of gamma and log gamma, beta and log beta
distribution implied in the expection step}

\label{an:Moments}

\subsection{First and second moments for the sufficient statistics of the
gamma \emph{pdf}}

Let $Z$ be a random variable with gamma distribution,
$Z\sim\Gamma(s,t)$. Using laplace transform it is easy to obtain
the first moment of $\ln(Z)$~:
\[
\mathbb{E}\left(  e^{\lambda\ln(Z)}\right)  =\mathbb{E}\left(
Z^{\lambda
}\right)  =\frac{t^{s}}{\Gamma(s)}\int_{0}^{+\infty}y^{\lambda}y^{s-1}%
e^{-ty}dy=\frac{\Gamma(s+\lambda)}{\Gamma(s)t^{\lambda}}.
\]
Differentiating this equation with respect to $\lambda$, we have
the expected value of $\ln(Z)$ (when $\lambda=0)$ and
$Z\ln(Z)$~(when $\lambda=1)$:
\begin{equation}
\left.  \frac{\partial\mathbb{E}\left(  Z^{\lambda}\right)
}{\partial\lambda }\right\vert _{\lambda=0}=\mathbb{E}\left(
\ln(Z)\right)  =\psi(s)-\ln(t),
\label{eq:FirstMoment}%
\end{equation}
and%
\begin{equation}
\quad\left.  \frac{\partial\mathbb{E}\left(  Z^{\lambda}\right)  }%
{\partial\lambda}\right\vert _{\lambda=1}=\mathbb{E}\left(
Z\ln(Z)\right)
=\frac{s}{t}\left(  \psi(s+1)-\ln(t)\right)  . \label{eq:ProdMoment}%
\end{equation}
Taking the second order derivative, we show~:%

\begin{equation}
\left.  \frac{\partial^{2}\mathbb{E}\left(  Z^{\lambda}\right)  }%
{\partial\lambda^{2}}\right\vert _{\lambda=0}=\mathbb{E}\left(
\ln (Z)^{2}\right)
=\psi^{\prime}(s)+\psi(s)^{2}-2\ln(t)\psi(s)+\ln(t)^{2}.
\label{eq:SecondMoment}%
\end{equation}

Therefore the variance-covariance matrix between $Z$ and $\ln(Z)$ is~:%
\[
\left(
\begin{array}
[c]{cc}%
\frac{s}{t^{2}} & \frac{1}{t}\\
\frac{1}{t} & \psi^{\prime}(s)
\end{array}
\right)
\]

\subsection{First and second moments for the sufficient statistics of the beta \emph{pdf}}

Let $S$ be a random variable with beta distribution
$S\sim\beta(s,t)$.
\[
\mathbb{E}\left(  e^{\lambda\ln(S)}\right)
=\frac{\Gamma(s+t)}{\Gamma
(s+t+\lambda)}\frac{\Gamma(s+\lambda)}{\Gamma(s)}%
\]
So that, by first and second differentiation, one gets,\ (the
derivation is quite straightfully performed if working with
$\ln\mathbb{E}\left( e^{\lambda\ln(S)}\right)  $)~:
\begin{align*}
\mathbb{E}\left(  \ln(S)\right)   &  =\psi(s)-\psi(s+t)\quad,\quad
\mathbb{E}\left(  \ln(1-S)\right)  =\psi(t)-\psi(s+t)\\
\mathbb{E}\left(  \ln(S)^{2}\right)   &
=\psi^{\prime}(s)-\psi^{\prime
}(s+t)+\left(  \psi(s)-\psi(s+t)\right)  ^{2}%
\end{align*}

One can extend the properties of characteristic function by
considering the
function of the two arguments $\lambda$ and $\mu$%
\[
\mathbb{E}\left(  e^{\lambda\ln(S)+\mu\ln(1-S)}\right)
=\frac{\Gamma
(s+t)}{\Gamma(s+t+\lambda)}\frac{\Gamma(s+\lambda)}{\Gamma(s)}\frac
{\Gamma(t+\mu)}{\Gamma(t)}%
\]

By cross-differentiation under regularity conditions (working with
$\ln\mathbb{E}\left(  S^{\lambda}(1-S)^{\mu}\right)  $ makes
things easier
here also) , the joint moment can be analytically obtained~:%

\begin{align*}
\left.  \frac{\partial^{2}\mathbb{E}\left(
S^{\lambda}(1-S)^{\mu}\right)
}{\partial\lambda\partial\mu}\right\vert _{\lambda=0,\mu=0}  &  =\mathbb{E}%
\left(  \ln(S)\ln(1-S)\right) \\
&  =-\psi^{\prime}(s+t)+\mathbb{E}\left(  \ln(S)\right)
\mathbb{E}\left( \ln(1-S)\right)
\end{align*}

Therefore the variance-covariance matrix between $\ln(S)$ and
$\ln(1-S)$ reads
:%
\[
\left(
\begin{array}
[c]{cc}%
\psi^{\prime}(s)-\psi^{\prime}(s+t) & -\psi^{\prime}(s+t)\\
-\psi^{\prime}(s+t) & \psi^{\prime}(t)-\psi^{\prime}(s+t)
\end{array}
\right)
\]

\section{EM algorithm principle}

\label{an:EM} From a constructive point of view, one often writes
\[
\lbrack x,z\left|  \theta\right.  ]=\left[  x\left|  \theta,z\right.  \right]
\times\left[  z\left|  \theta\right.  \right]  ,
\]
but using Bayes rule, we may write the reverse logarithmic form~:
\begin{equation}
\ln\left[  x\left|  \theta\right.  \right]  =\ln[x,z\left|  \theta\right.
]-\ln\left[  z\left|  \theta,x\right.  \right]  \label{eq:logvraisem}%
\end{equation}

Let us remark that relation \ref{eq:logvraisem} is valid whatever $z$ represents.

\subsection{Recall about EM algorithm and control of the gradient}

Under regularity conditions for the joint distribution $[x,z\left\vert
\theta\right.  ]$ and the conditional one $\left[  z\left\vert \theta
,x\right.  \right]  $ , integrating relation \ref{eq:logvraisem} with respect
to the probability density $\left[  z\left\vert \theta^{\prime},x\right.
\right]  $~:
\begin{align}
\ln\left[  x\left\vert \theta\right.  \right]   &  =\int\limits_{z}%
\ln[x,z\left\vert \theta\right.  ]\left[  z\left\vert \theta^{\prime
},x\right.  \right]  dz-\int\limits_{z}\ln\left[  z\left\vert \theta,x\right.
\right]  \left[  z\left\vert \theta^{\prime},x\right.  \right]  dz\nonumber\\
&  =Q(\theta,\theta^{\prime})-H(\theta,\theta^{\prime}) \label{eq:baseEM}%
\end{align}

The maximum of $\theta\mapsto H(\theta,\theta^{\prime})$ is achieved
in $\theta=\theta^{\prime}$ \citep{Tanner92}.

So $H(\theta,\theta^{\prime})<H(\theta^{\prime},\theta^{\prime})$.Let us
consider \ref{eq:baseEM} for $\theta$ and $\theta^{\prime}$
\[
\ln\left[  x\left|  \theta\right.  \right]  -\ln\left[  x\left|
\theta^{\prime}\right.  \right]  =\left(  Q(\theta,\theta^{\prime}%
)-Q(\theta^{\prime},\theta^{\prime})\right)  +\left(  H(\theta^{\prime}%
,\theta^{\prime})-H(\theta,\theta^{\prime})\right)
\]
EM algorithm is based upon an iterative procedure which exhibits $\theta$ such
that $Q(\theta,\theta^{\prime})>Q(\theta^{\prime},\theta^{\prime})$ . The best
$\theta$ is obtained by
\[
\theta=\underset{\theta}{argmax\ }Q(\theta,\theta^{\prime})
\]

During iteration we can monitor the value of the gradient for the log
likelihood~:
\begin{equation}
\frac{\partial\ln\left[  x\left\vert \theta\right.  \right]  }{\partial\theta
}=\frac{\partial\ln[x,z\left\vert \theta\right.  ]}{\partial\theta}%
-\frac{\partial\ln\left[  z\left\vert \theta,x\right.  \right]  }%
{\partial\theta} \label{eq:gradientlogvrais}%
\end{equation}

Integrating the right hand term with respect to conditional
density $\left[ z\left|  \theta,x\right.  \right]  ,$and keeping
in mind that, for any sufficiently regular \emph{pdf}
$f(z;\theta)$ of variable $z$ with parameter
$\theta$ one can write: $%
{\displaystyle\int\limits_{z}}
\frac{\partial\ln f(z,\theta)}{\partial\theta}f(z;\theta)dz=$ $\frac{\partial
}{\partial\theta}%
{\displaystyle\int\limits_{z}}
\frac{\partial\ln f(z,\theta)}{\partial\theta}f(z;\theta)dz=0,$ we have
\begin{align}
\frac{\partial\ln\left[  x\left|  \theta\right.  \right]  }{\partial\theta}
&  =%
{\displaystyle\int\limits_{z}}
\frac{\partial\ln[x,z\left|  \theta\right.  ]}{\partial\theta}\left[  z\left|
\theta,x\right.  \right]  dz-%
{\displaystyle\int\limits_{z}}
\frac{\partial\ln\left[  z\left|  \theta,x\right.  \right]  }{\partial\theta
}\left[  z\left|  \theta,x\right.  \right]  dz\nonumber\\
\frac{\partial\ln\left[  x\left|  \theta\right.  \right]  }{\partial\theta}
&  =%
{\displaystyle\int\limits_{z}}
\frac{\partial\ln[x,z\left|  \theta\right.  ]}{\partial\theta}\left[  z\left|
\theta,x\right.  \right]  dz \label{eq:condexpscore}%
\end{align}

We may use this equality (computed by Monte Carlo method) to perform a
gradient method to obtain the maximum likelihood or just to check along the
iterations that the gradient is going to zero.

\subsection{Score function}

From now on, let's call $Sc(\theta,z,x)$ = $\frac{\partial\ln[x,z\left\vert
\theta\right.  ]}{\partial\theta}$the score, i.e the complete loglikelihood
gradient and $Sc(\theta_{i},z,x)$ = $\frac{\partial\ln[x,z\left\vert
\theta\right.  ]}{\partial\theta_{i}}$ its $i^{th}$ component. $\nabla\theta,$
equation \ref{eq:condexpscore} proves that its conditional expectation (with
respect to $\left[  z\left\vert \theta,x\right.  \right]  $) is always equal
to the likelihood gradient. Pushing the derivation game one step further leads
to:%
\begin{align}
\frac{\partial}{\partial\theta_{j}}\left\{  \frac{\partial\ln\left[
x\left\vert \theta\right.  \right]  }{\partial\theta_{i}}\right\}   &  =%
{\displaystyle\int\limits_{z}}
\left\{  \frac{\partial Sc_{i}}{\partial\theta_{j}}\left[  z\left\vert
\theta,x\right.  \right]  +Sc_{i}\frac{\partial\left[  z\left\vert
\theta,x\right.  \right]  }{\partial\theta_{j}}\frac{\left[  z\left\vert
\theta,x\right.  \right]  }{\left[  z\left\vert \theta,x\right.  \right]
}\right\}  dz\nonumber\\
\frac{\partial^{2}\ln\left[  x\left\vert \theta\right.  \right]  }%
{\partial\theta_{i}\partial\theta_{j}}  &  =%
{\displaystyle\int\limits_{z}}
\left\{  \frac{\partial^{2}\ln\left[  x,z\left\vert \theta\right.  \right]
}{\partial\theta_{i}\partial\theta_{j}}+Sc_{i}\left(  Sc_{j}-\frac{\partial
\ln\left[  x\left\vert \theta\right.  \right]  }{\partial\theta_{j}}\right)
\right\}  \left[  z\left\vert \theta,x\right.  \right]  dz
\label{eq:derivscore}%
\end{align}

\subsection{Information matrix}

\label{an:Infomatrix} To obtain the covariance matrix of the estimators at the
maximum of likelihood, the empirical information matrix needs to be computed.
The second order derivative is obtained by differentiating
\ref{eq:logvraisem}:%
\begin{equation}
\frac{\partial^{2}\ln\left[  x\left|  \theta\right.  \right]  }{\partial
\theta_{i}\partial\theta_{j}}=\frac{\partial^{2}\ln[x,z\left|  \theta\right.
]}{\partial\theta_{i}\partial\theta_{j}}-\frac{\partial^{2}\ln\left[  z\left|
\theta,x\right.  \right]  }{\partial\theta_{i}\partial\theta_{j}}
\label{eq:derive2_un}%
\end{equation}

At the maximum $\theta=\hat{\theta}$, formula \ref{eq:gradientlogvrais}
implies$\frac{\partial\ln\left[  x\left\vert \theta\right.  \right]
}{\partial\theta_{jj}}=0$ so that equation \ref{eq:derivscore} takes a more
friendly aspect because the score term $\frac{\partial\ln\left[  x\left\vert
\theta\right.  \right]  }{\partial\theta_{j}}$in the right hand side vanishes
at $\theta=\hat{\theta}$ . Equation \ref{eq:derive2_un} becomes~therefore much
more handy because it only involves conditional expectations of first and
second derivatives of the complete likelihood terms :
\begin{equation}
\frac{\partial^{2}\ln(\left[  x\left\vert \hat{\theta}\right.  \right]
)}{\partial\theta_{i}\partial\theta_{j}}=\int\limits_{z}\left(  \frac
{\partial^{2}\ln\left[  x,z\left\vert \hat{\theta}\right.  \right]  }%
{\partial\theta_{i}\partial\theta_{j}}+\frac{\partial\ln\left[  x,z\left\vert
\hat{\theta}\right.  \right]  }{\partial\theta_{i}}\frac{\partial\ln\left[
x,z\left\vert \hat{\theta}\right.  \right]  }{\partial\theta_{j}}\right)
\left[  z\left\vert \hat{\theta},x\right.  \right]  dz \label{eq:2dderivative}%
\end{equation}

As $\int\limits_{z}\left(  \frac{\partial\ln\left[  x,z\left\vert \hat{\theta
}\right.  \right]  }{\partial\theta_{j}}\right)  \left[  z\left\vert
\hat{\theta},x\right.  \right]  dz=\frac{\partial\ln\left[  x\left\vert
\hat{\theta}\right.  \right]  }{\partial\theta_{j}}=0,$ the second term in the
right hand side of eq \ref{eq:2dderivative} can be considered as the
conditional variance of the gradient of the complete log-likelihood
$\ln\left[  x,z\left\vert \hat{\theta}\right.  \right]  $ . This expectation
can be numerically computed with the same techniques to which recourse was
made for the EM algorithm.

\section{Detailed proofs of  propositions}
\label{subsec:proof}
\subsection{Proof of proposition \ref{prop:murhocond}}
\label{subsec:proof1}
Since we detail the computation for one
particular $s$, we will omit to mention it in order to make the
reading easier. We also note respectively $\underline{y}$,
$\underline{D}$ and $\underline{N}$ the vectors of data, catching
efforts and corresponding number of clumps in one stratum.\newline
We define $J$ as
\begin{equation}
J(\underline{N},\rho,\mu) =\left[  \rho,\mu,\underline{N}\left\vert
a,b,c,d,\underline{y},\underline{D}\right.  \right]  .
\end{equation}
Then $J$ satisfies the following set of equations~:
\begin{align*}
&  \varpropto\left[  \underline{y},\rho,\mu,\underline{N}\left\vert a,b,c ,d
,\underline{D}\right.  \right] \\
&  \varpropto\left(  \prod\limits_{i=1}^{I}\left[  y_{i}\left\vert N_{i}%
,\rho\right.  \right]  \left[  N_{i}\left\vert \mu,D_{i}\right.  \right]
\right)  \left[  \mu\left\vert a ,b \right.  \right]  \left[  \rho\left\vert c
,d \right.  \right] \\
&  \varpropto\left(  \prod\limits_{i=1}^{I}\left[  y_{i}\left\vert N_{i}%
,\rho,\mu\right.  \right]  \left[  N_{i}\left\vert \rho,\mu\right.  \right]
\right)  \left(  \mu^{a -1}e^{-\mu b }\right)  \left(  \rho^{c -1}e^{-\rho d
}\right)
\end{align*}
with the convention that $[A|B]\varpropto f(A,B)$ means that the coefficient
of proportionality only depends on $B$. \medskip We note $I^{\star}$ the
number of zero value $y$ and we reorder the vector $y$ so that the
$I^{+}=I-I^{\ast}$ non zero $y_{i}$ are the first, so that $J$ may be written
as~: {\small
\begin{align*}
J(\underline{N},\rho,\mu)  &  \varpropto\left(  \prod\limits_{i=1}%
^{I-I^{\star}}\left(  y_{i}^{N_{i}}e^{-\rho y_{i}}\frac{\rho^{N_{i}}}%
{\Gamma(N_{i})}\right)  \left(  \frac{e^{-\mu D_{i}}(\mu\,D_{i})^{N_{i}}%
}{\Gamma(N_{i}+1)}\right)  \right) \\
&  \hspace{1cm}\left(  \prod\limits_{i^{\star}=I-I^{\star}+1}^{I}%
\delta(N_{i^{\ast}})e^{-\mu D_{i^{\star}}}\right)  \left(  \mu^{a -1}e^{-\mu
b}\right)  \left(  \rho^{c -1}e^{-\rho d}\right)
\end{align*}
} Defining $Y_{+}=\sum_{i=1}^{I}y_{i}$, $N_{+}=\sum_{i=1}^{I}N_{i}$ and
$D_{+}=\sum_{i=1}^{I}D_{i}$, we obtain~:
\[
J(\underline{N},\rho,\mu)\varpropto\left(  \prod_{i=1}^{I^{+}}\displaystyle
\frac{y_{i}^{N_{i}}}{\Gamma(N_{i})\Gamma(N_{i}+1)}\right)  e^{-\rho(Y_{+}+d
)}\rho^{N_{+}+c -1}e^{-\mu(D_{+}+b )}\mu^{N_{+}+a -1}%
\]
Conditionally to the latent vector $\underline{N}$, the random effects $\rho$
and $\mu$ are independent. Isolating the terms which depend on $\mu$ on one
side and those depend on $\rho$ on the other, we find that
\begin{align*}
\left[  \mu|\underline{N},\theta,\underline{y},\underline{D}\right]   &
\sim\Gamma(a +N_{+},b +D_{+})\\
\left[  \rho|\underline{N},\theta,\underline{y}\right]   &  \sim\Gamma(c
+N_{+},d +Y_{+})
\end{align*}
For the expectation step we only need to compute
$\mathbb{E}_{\theta}\left( \mu_{s}\mid\underline{Y_{s}}\right)  $,
$\mathbb{E}_{\theta}\left(  \ln
(\mu_{s})\mid\underline{Y_{s}}\right)  $ and the same sufficient
statistics concerning $\rho$.\newline Since
$\mu_{s}|\underline{N_{s}},\theta ,\underline{y}$ follows a gamma
distribution $\Gamma(a+N_{s+},b+D_{s+})$, the conditional expected
value $\mu_{s}$ given $\underline{N_{s}}$ and $\theta=(a,b,c,d)$ is
$(a+N_{s+})/(b+D_{s+})$. \newline Then
\[
\mathbb{E}_{\theta^{\prime}}\left(  \mu_{s}\mid\underline{y_{s}}\right)
=\mathbb{E}_{\theta^{\prime}}\left(  \frac{a^{\prime}+N_{s+}}{b^{\prime
}+D_{s+}}\left\vert \underline{y}\right.  \right)  =\frac{a^{\prime
}+\mathbb{E}_{\theta^{\prime}}\left(  N_{s+}|\underline{y_{s}}\right)
}{b^{\prime}+D_{s+}}.
\]
If $Z$ follows gamma distribution $\Gamma(s,t)$, then $\mathbb{E}(\ln
(Z))=\psi(s)-\ln(t)$ (see annex \ref{an:Moments}), so that
\[
\mathbb{E}_{\theta^{\prime}}\left(  \ln(\mu_{s})\mid\underline{y_{s}}\right)
=\mathbb{E}_{\theta^{\prime}}\left(  \psi(a^{\prime}+N_{s+})\left\vert
\underline{y_{s}}\right.  \right)  -\ln(b^{\prime}+D_{s+}).
\]
We have respectively for $\rho_{s}$
\[
\mathbb{E}_{\theta^{\prime}}\left(  \rho_{s}\mid\underline{y_{s}}\right)
=\frac{c^{\prime}+\mathbb{E}_{\theta^{\prime}}\left(  N_{s+}|\underline{y_{s}%
}\right)  }{d^{\prime}+Y_{s+}},
\]
and
\[
\mathbb{E}_{\theta^{\prime}}\left(  \ln(\rho_{s})\mid\underline{y_{s}}\right)
=\mathbb{E}_{\theta^{\prime}}\left(  \psi(c^{\prime}+N_{s+})\left\vert
\underline{y_{s}}\right.  \right)  -\ln(d^{\prime}+Y_{s+}).
\]
%
%

\subsection{Proof of proposition \ref{prop:Ncond}}

Let us define  $J$ as the distribution of  $\left[
\rho,\mu,\underline{N}\left\vert \theta^{\prime},
\underline{y},\underline{D}\right.  \right]$ in one particular
stratum $s$. We will write $J$ in a bottom-up perspective and
consider the distribution of $\mu$ and $\rho$  conditionned  by $N$,
because $\mu$ and $\rho$ are conditionally independant.

$J$ is given by~:
\begin{align*}
J(\underline{N},\rho,\mu)  &  = \left[  \rho,\mu,\underline{N}\left\vert
\theta^{\prime}, \underline{y},\underline{D}\right.  \right]  \\
&  = \left[  \rho\left\vert \underline{N}, \theta,\underline{y}\right.
\right]  \left[  \mu\left\vert \underline{N}, \theta^{\prime},\underline
{y},\underline{D}\right.  \right]  \left[  \underline{N} \left\vert ,
\theta,\underline{y},\underline{D}\right.  \right]
\end{align*}

Using the independent conditional gamma distributions of $\mu$ and $\rho$ and
integrating according to $\mu$ and $\rho$ given $\underline{N},$ we can
exhibit all the terms depending on $\underline{N}$.
\begin{align*}
\int_{\rho}\int_{\mu}  &  J(\underline{N},\rho,\mu)d\mu\,d\rho=\left[
\underline{N}\left\vert \theta,\underline{y},\underline{D}\right.  \right] \\
&  \propto\prod_{i=1}^{I^{+}}\left(  \frac{y_{i}^{N_{i}}}{\Gamma(N_{i}%
)\Gamma(N_{i}+1)}\right)  \prod_{i^{\star}=I-I^{\star+1}}^{I}\delta
(N_{i^{\star}})\left(  \frac{(b^{\prime}+D_{+})^{N_{+}}(d+Y_{+})^{N_{+}}%
}{\Gamma(a+N_{+})\Gamma(c+N_{+})}\right)  ^{-1}%
\end{align*}

\subsection{Proof of proposition \ref{prop:Fisher}}

\begin{figure}[tbhptbh]
\begin{center}
\includegraphics[
trim=0.000000in 0.000000in -1.367203in 0.000000in,
natheight=6.146200in, natwidth=3.531000in, height=5.0215cm,
width=4.0154cm
]{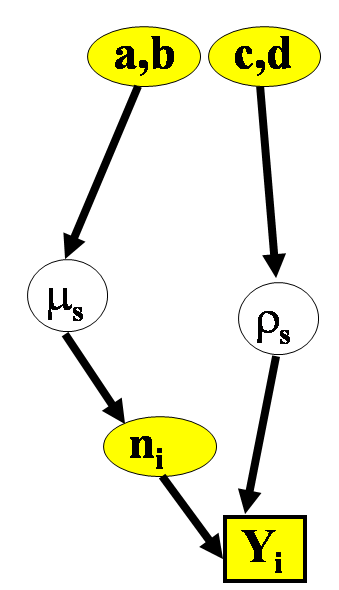}
\end{center}
\caption{Given $N,\rho_{s}\perp\mu_{s}$}%
\label{Fig:muro}%
\end{figure}

%
%
In the following $Z$ will stand for all the hidden variables i.e
$\underline{\mathbf{Z}}=\left(  \underline{\mathbf{N}},\bm{\mu},\bm{\rho
}\right)  $ , $\left\vert M_{ij}\right\vert $ is another notation for matrix
$M$ that details the content of the $i^{th}$ row and $j^{th}$ column,
and\ $\frac{\partial F(\theta)}{\partial\theta}$ stands for the gradient of
$F$ written as a vector whose $i^{th}$ component is the scalar $\frac{\partial
F(\theta)}{\partial\theta_{i}}$ . The key equation involves \ rewriting
equation\ \ref{eq:2dderivative} as the expectation of the second order
derivative of the complete log-likelihood and the variance of the score (its
gradient) to be taken with regards to the conditional distribution $\left[
Z\left\vert x,\hat{\theta}\right.  \right]  $ (see annex \ref{an:Infomatrix})%
\begin{equation}
\left\vert \frac{\partial^{2}\ln(\left[  x\left\vert \hat{\theta}\right.
\right]  )}{\partial\theta_{i}\partial\theta_{j}}\right\vert =\mathbb{E}%
_{Z|x}\left\vert \frac{\partial^{2}\ln(\left[  x,Z\left\vert \hat{\theta
}\right.  \right]  )}{\partial\theta_{i}\partial\theta_{j}}\right\vert
+\Var_{Z|x}\left(  \frac{\partial\ln\left[  x,Z\left\vert
\theta\right.  \right]  }{\partial\theta}\right)  \label{key-vrais}%
\end{equation}
Computing the first term of the right hand side of equation \ref{key-vrais} is
easy, since $\left[  x\left\vert z,\theta\right.  \right]  =\left[
x\left\vert z\right.  \right]  $ (consequently the complete log-likelihood
$\ln(\left[  x,z\left\vert \theta\right.  \right]  )$ can be separated as
$\ln(\left[  x\left\vert z\right.  \right]  )+\ln(\left[  z,\left\vert
\theta\right.  \right]  )$) and the gamma random effects $\left[
z|\theta\right]  $ belong to an exponential family. As a consequence, annex
\ref{an:secondderiv}\ shows that
\[
\mathbb{E}\left\vert \frac{\partial^{2}\ln(\left[  x,Z\left\vert \hat{\theta
}\right.  \right]  )}{\partial\theta_{i}\partial\theta_{j}}\right\vert
=\left\vert \frac{\partial^{2}\ln(\left[  x,z\left\vert \hat{\theta}\right.
\right]  )}{\partial\theta_{i}\partial\theta_{j}}\right\vert =S\,\left(
\begin{array}
[c]{cccc}%
-\psi^{\prime}(\hat{a}) & \frac{1}{\hat{b}} & 0 & 0\\
\frac{1}{b} & -\frac{\hat{a}}{\hat{b}^{2}} & 0 & 0\\
0 & 0 & -\psi^{\prime}(\hat{c}) & \frac{1}{\hat{d}}\\
0 & 0 & \frac{1}{\hat{d}} & \frac{-\hat{c}}{\hat{d}^{2}}%
\end{array}
\right)
\]
\begin{figure}[tbh]
\begin{center}
\includegraphics[
trim=0.000000in 0.000000in 0.000000in -0.484474in,
natheight=7.499600in, natwidth=9.999800in, height=8.0309cm,
width=10.0364cm
]{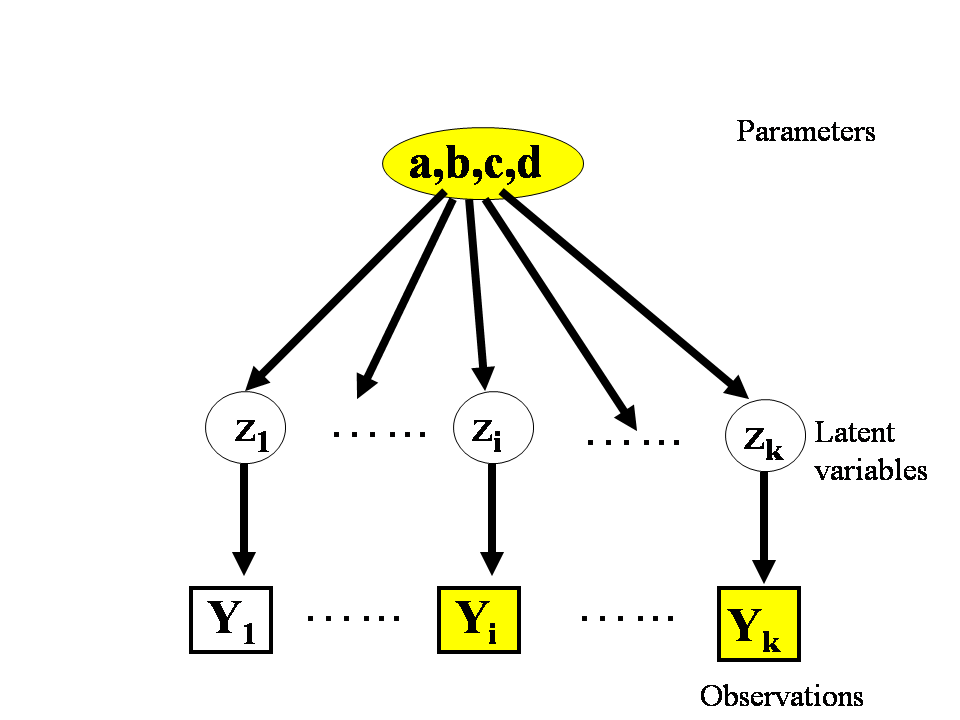}
\end{center}
\caption{The random effects in each stratum are conditionally
independent given
the data and the set of parameters}%
\label{Fig:hierach}%
\end{figure}As shown in Figure~\ref{Fig:hierach}. , given $Y_{s},Y_{s^{\prime
}}$ and $\theta,$ the latent variables $Z_{s}$ and $Z_{s^{\prime}}$ of two
stratum $s$ and $s^{\prime}$ are conditionnaly independent, therefore~:%
\[
\Var_{Z|x}\left(  \frac{\partial\ln\left[  x,Z\left\vert
\theta\right.  \right]  }{\partial\theta}\right)  =\sum_{s=1}^{S}%
\Var_{Z_{s}|x}\left(
\begin{array}
[c]{c}%
\ln(\mu_{s})\\
-\mu_{s}\\
\ln(\rho_{s})\\
-\rho_{s}%
\end{array}
\right)
\]
To evaluate the variance of the score in stratum s, we will take
advantage of successive conditioning due to the hierarchical
structure depicted in Figure~\ref{Fig:DagRlol}. Recalling that the
latent variable $Z_{s}$ includes, in addition to
$(\mu_{s},\rho_{s})$, the vector $\underline{N_{s}}$ , i-e the
latent number of clumps for each record, the variance conditional
decomposition formula gives:%
\[
\Var_{Z_{s}|x}\left(
\begin{array}
[c]{c}%
\ln(\mu_{s})\\
-\mu_{s}\\
\ln(\rho_{s})\\
-\rho_{s}%
\end{array}
\right)  =\mathbb{E}_{\underline{N_{s}}|x}\left(  \Var\left(
\begin{array}
[c]{c}%
\ln(\mu_{s})\\
-\mu_{s}\\
\ln(\rho_{s})\\
-\rho_{s}%
\end{array}
|\underline{N_{s}}\right)  \right)  +\Var_{\underline{N_{s}}|x}\left(
\mathbb{E}\left(
\begin{array}
[c]{c}%
\ln(\mu_{s})\\
-\mu_{s}\\
\ln(\rho_{s})\\
-\rho_{s}%
\end{array}
|\underline{N_{s}}\right)  \right)
\]
So that we have
\[
I_{e}(\hat{\theta},x)=S\,\left(
\begin{matrix}
-\psi^{\prime}(\hat{a}) & \frac{1}{\hat{b}} & 0 & 0\\
\frac{1}{\hat{b}} & -\frac{\hat{a}}{\hat{b}^{2}} & 0 & 0\\
0 & 0 & -\psi^{\prime}(\hat{c}) & \frac{1}{\hat{d}}\\
0 & 0 & \frac{1}{\hat{d}} & -\frac{\hat{c}}{\hat{d}^{2}}\\
&  &  &
\end{matrix}
\right)  +\sum_{s=1}^{S}(A_{s}+B_{s})
\]
with
\[
A_{s}=\mathbb{E}_{\underline{\mathbf{N}}|x}\left(  \Var\left(
\begin{array}
[c]{c}%
\ln(\mu_{s})\\
-\mu_{s}\\
\ln(\rho_{s})\\
-\rho_{s}%
\end{array}
|\underline{\mathbf{N}}\right)  \right)  \quad\mbox{and}\quad B_{s}%
=\Var_{\underline{N_{s}}|x}\left(  \mathbb{E}\left(
\begin{array}
[c]{c}%
\ln(\mu_{s})\\
-\mu_{s}\\
\ln(\rho_{s})\\
-\rho_{s}%
\end{array}
|\underline{N_{s}}\right)  \right)  .
\]
Given $\underline{N_{s}}$, $\mu_{s}$ and $\rho_{s}$ are
independent. Moreover the \emph{pdf}
$[\rho_{s}|\underline{N_{s}},\underline{Y_{s}},a,b,c,d]$ and $[\mu
_{s}|\underline{N_{s}},\underline{Y_{s}},a,b,c,d]$ are gamma and
analytic expressions are available for the expectation and
variance of the gamma sufficient statistics, as detailed in
equations \ref{eq:FirstMoment} to \ref{eq:SecondMoment}. The key
functions of $N_{s+}$ are ($a_{s}\prime
,b_{s}\prime,c_{s}\prime,d_{s}\prime)=(a+N_{s+},b+D_{s+},c+N_{s+},d+Y_{s+})$
such that~:%
\[
\mathbb{E}\left(
\begin{array}
[c]{c}%
\ln(\mu_{s})\\
-\mu_{s}\\
\ln(\rho_{s})\\
-\rho_{s}%
\end{array}
|\underline{\mathbf{N}}\right)  =\left(
\begin{array}
[c]{c}%
\psi(a_{s}^{\prime})-\ln(b_{s}^{\prime})\\
-\frac{a_{s}^{\prime}}{b_{s}^{\prime}}\\
\psi(c_{s}^{\prime})-\ln(d_{s}^{\prime})\\
-\frac{c_{s}^{\prime}}{d_{s}^{\prime}}%
\end{array}
\right)
\]
and then $B_{s}$ is obtained by taking the covariance of this vector~:
\[
B_{s}=\Var_{N_{s+}|\hat{\theta},x}\left(  \mathbb{E}\left(
\begin{array}
[c]{c}%
\ln(\mu_{s})\\
-\mu_{s}\\
\ln(\rho_{s})\\
-\rho_{s}%
\end{array}
|\underline{\mathbf{N_{s+}}}\right)  \right)
\]
Given $\underline{N_{s}}$ additional advantage is taken from the
conditional independence of $\rho_{s}$ and $\mu_{s}$ as shown in
Figure~\ref{Fig:muro}, .
\[
\Var\left(
\begin{array}
[c]{c}%
\ln(\mu_{s})\\
-\mu_{s}\\
\ln(\rho_{s})\\
-\rho_{s}%
\end{array}
|\underline{\mathbf{N}}\right)  =\left(
\begin{array}
[c]{cccc}%
-\psi^{\prime}(a^{\prime}) & \frac{1}{b^{\prime}} & 0 & 0\\
\frac{1}{b^{\prime}} & -\frac{a^{\prime}}{b^{\prime2}} & 0 & 0\\
0 & 0 & -\psi^{\prime}(c^{\prime}) & \frac{1}{d^{\prime}}\\
0 & 0 & \frac{1}{d^{\prime}} & \frac{-c^{\prime}}{d^{\prime2}}%
\end{array}
\right)
\]
and the expression for $A_{s}$ follows easily.
%
%

\section{Second derivative of the complete log-likelihood}

\label{an:secondderiv} Let us first recall the complete log likelihood of the
model~:
\begin{align*}
\ln\left[  x,z\left\vert \theta\right.  \right]   &  =C_{-\theta}%
+(a-1)\sum\limits_{s=1}^{S}\ln\mu_{s}+Sa\ln b-b\sum\limits_{s=1}^{S}\mu
_{s}-S\ln\Gamma(a)\\
&  +(c-1)\sum\limits_{s=1}^{S}\ln\rho_{s}+Sc\ln d-d\sum\limits_{s=1}^{S}%
\rho_{s}-S\ln\Gamma(c)
\end{align*}

In the first derivative, the latent variables $\bm{\mu}$ and
$\bm{\rho}$ appear not surprisingly only through their arithmetic or
geometric means (sufficient statistics for the gamma \emph{pdf}).
Using standard notation $\bar{\mu}$ for the arithmetic mean
$\frac{1}{S}\sum_{s=1}^{S}\mu_{s}$, we have~:
\begin{align*}
\frac{\partial\ln\left[  x,z\left|  \theta\right.  \right]  }{\partial
a}=S\left(  \overline{\ln(\mu)}+\ln b-\psi(a)\right)   &  \qquad\frac
{\partial\ln\left[  x,z\left|  \theta\right.  \right]  }{\partial c}=S\left(
\overline{\ln(\rho)}+\ln d-\psi(c)\right) \\
\frac{\partial\ln\left[  x,z\left|  \theta\right.  \right]  }{\partial
b}=S\left(  \frac{a}{b}-\overline{\mu}\right)   &  \qquad\frac{\partial
\ln\left[  x,z\left|  \theta\right.  \right]  }{\partial d}=S\left(  \frac
{c}{d}-\overline{\rho}\right)
\end{align*}
The gradient of the complete log-likelihood \ (so-called the ''score'') may be
split into two parts~: the first one $\Delta_{\theta}$ does not depend on the
latent variable $z$ while the other one $\Delta_{z}$ gathers terms depending
on $z$ (and possibly of $\theta$), i.e~:%

\[
\left(  \frac{\partial\ln\left[  x,z\left\vert \hat{\theta}\right.  \right]
}{\partial\theta}\right)  =\Delta_{\theta}+\Delta_{z}%
\]

with
\[%
\begin{array}
[c]{cc}%
\Delta_{\theta} =S \left(
\begin{array}
[c]{c}%
\ln b-\psi(a)\\
\frac{a}{b}\\
\ln d-\psi(c)\\
\frac{c}{d}%
\end{array}
\right)  & \Delta_{z} = S \left(
\begin{array}
[c]{c}%
\overline{\ln(\mu)}\\
-\overline{\mu}\\
\overline{\ln(\rho)}\\
-\overline{\rho}%
\end{array}
\right)
\end{array}
\]

In addition here, $\Delta_{z}$ does not contain terms with $\theta$,
consequently the second order derivatives are easy to obtain and
don't involve the latent variable~:
\begin{align*}
\frac{\partial^{2}\ln\left[  x,z\left|  \theta\right.  \right]  }{\partial
a\partial a}=-S\psi^{\prime}(a)  &  \quad\frac{\partial^{2}\ln\left[
x,z\left|  \theta\right.  \right]  }{\partial c\partial c}=-S\psi^{\prime
}(c)\\
\frac{\partial^{2}\ln\left[  x,z\left|  \theta\right.  \right]  }{\partial
a\partial b}=\frac{S}{b}  &  \quad\frac{\partial^{2}\ln\left[  x,z\left|
\theta\right.  \right]  }{\partial c\partial d}=\frac{S}{d}\\
\frac{\partial^{2}\ln\left[  x,z\left|  \theta\right.  \right]  }{\partial
b\partial b}=-\frac{Sa}{b^{2}}  &  \quad\frac{\partial^{2}\ln\left[
x,z\left|  \theta\right.  \right]  }{\partial d\partial d}=\frac{-Sc}{d^{2}}%
\end{align*}

\section{Second derivative of the complete log-likelihood with discrete data}

\label{an:secondderivdiscrete} The complete log likelihood of the model, in
the discrete case, reads ~:
\begin{align*}
\ln\left[  x,z\left|  \theta\right.  \right]   &  =C_{-\theta}+(a-1)\sum
\limits_{s=1}^{S}\ln\mu_{s}+Sa\ln b-b\sum\limits_{s=1}^{S}\mu_{s}-S\ln
\Gamma(a)\\
S\ln\left(  \frac{\Gamma(c+d)}{\Gamma(c)\Gamma(d)}\right)   &  +(c-1)\sum
\limits_{s=1}^{S}\ln p_{s}+(d-1)\sum\limits_{s=1}^{S}\ln(1-p_{s})
\end{align*}

In the first derivative, the latent variables $\bm{\mu}$ and
$\bm{p}$ appear only through their arithmetic or geometric means
(sufficient statistics for the gamma and beta \emph{pdf}). Using
standard notation $\bar{\mu}$ for the arithmetic mean
$\frac{1}{S}\sum_{s=1}^{S}\mu_{s}$, we have~:
\begin{align*}
\frac{\partial\ln\left[  x,z\left|  \theta\right.  \right]  }{\partial
a}=S\left(  \overline{\ln(\mu)}+\ln b-\psi(a)\right)   &  \qquad\frac
{\partial\ln\left[  x,z\left|  \theta\right.  \right]  }{\partial c}=S\left(
\overline{\ln(p)}+\psi(c+d)-\psi(c)\right) \\
\frac{\partial\ln\left[  x,z\left|  \theta\right.  \right]  }{\partial
b}=S\left(  \frac{a}{b}-\overline{\mu}\right)   &  \qquad\frac{\partial
\ln\left[  x,z\left|  \theta\right.  \right]  }{\partial d}=S\left(
\overline{\ln(1-p)}+\psi(c+d)-\psi(d)\right)
\end{align*}
The gradient of the complete log-likelihood \ (so-called the ''score'') may be
split into two parts~: the first one $\Delta_{\theta}$ does not depend on the
latent variable $z$ while the other one $\Delta_{z}$ gathers terms depending
on $z$ (and possibly of $\theta$), i.e~:%

\[
\left(  \frac{\partial\ln\left[  x,z\left\vert \hat{\theta}\right.  \right]
}{\partial\theta}\right)  =\Delta_{\theta}+\Delta_{z}%
\]

with
\[%
\begin{array}
[c]{cc}%
\Delta_{\theta}=S\left(
\begin{array}
[c]{c}%
\ln b-\psi(a)\\
\frac{a}{b}\\
\psi(c+d)-\psi(c)\\
\psi(c+d)-\psi(d)
\end{array}
\right)  & \Delta_{z}=S\left(
\begin{array}
[c]{c}%
\overline{\ln(\mu)}\\
-\overline{\mu}\\
\overline{\ln(p)}\\
\overline{\ln(1-p)}%
\end{array}
\right)
\end{array}
\]

In addition here, $\Delta_{z}$ does not contain terms with $\theta$,
consequently the second order derivatives are easy to obtain and
don't involve the latent variable; with $Z$ standing for all the
hidden variables i.e $\underline{\mathbf{Z}}=\left(
\underline{\mathbf{N}},\bm{\mu},\bm {p}\right)  $:
\[
\left|  \frac{\partial^{2}\ln(\left[  x,z\left|  \hat{\theta}\right.  \right]
)}{\partial\theta_{i}\partial\theta_{j}}\right|  =S\,\left(
\begin{array}
[c]{cccc}%
-\psi^{\prime}(\hat{a}) & \frac{1}{\hat{b}} & 0 & 0\\
\frac{1}{b} & -\frac{\hat{a}}{\hat{b}^{2}} & 0 & 0\\
0 & 0 & -\psi^{\prime}(\hat{c})+\psi^{\prime}(\hat{c}+\hat{d}) & \psi^{\prime
}(\hat{c}+\hat{d})\\
0 & 0 & \psi^{\prime}(\hat{c}+\hat{d}) & -\psi^{\prime}(\hat{d})+\psi^{\prime
}(\hat{c}+\hat{d})
\end{array}
\right)
\]
As shown in Figure~\ref{Fig:hierach} for the continuous case , given
$Y_{s},Y_{s^{\prime}}$ and $\theta,$ the latent variables $Z_{s}$
and $Z_{s^{\prime}}$ of two strata $s$ and $s^{\prime}$ are
conditionnaly
independent, therefore~:%
\[
\Var_{Z|x}\left(  \frac{\partial\ln\left[  x,Z\left|  \theta\right.
\right]  }{\partial\theta}\right)  =\sum_{s=1}^{S}\Var_{Z_{s}%
|x}\left(
\begin{array}
[c]{c}%
\ln(\mu_{s})\\
-\mu_{s}\\
\ln(p_{s})\\
\ln(1-p_{s})
\end{array}
\right)
\]
To evaluate the variance of the score in stratum s, we will take
advantage from successive conditioning due to the hierarchical
structure depicted in Figure~\ref{Fig:DagRlol} still true for the
discrete case. The variance conditional
decomposition formula gives:%
\[
\Var_{Z_{s}|x}\left(
\begin{array}
[c]{c}%
\ln(\mu_{s})\\
-\mu_{s}\\
\ln(\rho_{s})\\
-\rho_{s}%
\end{array}
\right)  =\mathbb{E}_{\underline{N_{s}}|x}\left(  \Var\left(
\begin{array}
[c]{c}%
\ln(\mu_{s})\\
-\mu_{s}\\
\ln(p_{s})\\
\ln(1-p_{s})
\end{array}
|\underline{N_{s}}\right)  \right)  +\Var_{\underline{N_{s}}|x}\left(
\mathbb{E}\left(
\begin{array}
[c]{c}%
\ln(\mu_{s})\\
-\mu_{s}\\
\ln(p_{s})\\
\ln(1-p_{s})
\end{array}
|\underline{N_{s}}\right)  \right)
\]
So that we have
\[
I_{e}(\hat{\theta},x)=S\,\left(
\begin{array}
[c]{cccc}%
-\psi^{\prime}(\hat{a}) & \frac{1}{\hat{b}} & 0 & 0\\
\frac{1}{b} & -\frac{\hat{a}}{\hat{b}^{2}} & 0 & 0\\
0 & 0 & -\psi^{\prime}(\hat{c})+\psi^{\prime}(\hat{c}+\hat{d}) & \psi^{\prime
}(\hat{c}+\hat{d})\\
0 & 0 & \psi^{\prime}(\hat{c}+\hat{d}) & -\psi^{\prime}(\hat{d})+\psi^{\prime
}(\hat{c}+\hat{d})
\end{array}
\right)  +\sum_{s=1}^{S}(A_{s}+B_{s})
\]
with
\[
A_{s}=\mathbb{E}_{\underline{\mathbf{N}}|x}\left(  \Var\left(
\begin{array}
[c]{c}%
\ln(\mu_{s})\\
-\mu_{s}\\
\ln(p_{s})\\
\ln(1-p_{s})
\end{array}
|\underline{\mathbf{N}}\right)  \right)  \quad\mbox{and}\quad B_{s}%
=\Var_{\underline{N_{s}}|x}\left(  \mathbb{E}\left(
\begin{array}
[c]{c}%
\ln(\mu_{s})\\
-\mu_{s}\\
\ln(p_{s})\\
\ln(1-p_{s})
\end{array}
|\underline{N_{s}}\right)  \right)  .
\]
Given $\underline{N_{s}}$, $\mu_{s}$ and $\rho_{s}$ are
independent. Moreover the \emph{pdf}
$[\rho_{s}|\underline{N_{s}},\underline{Y_{s}},a,b,c,d]$ and
$[p_{s}|\underline{N_{s}},\underline{Y_{s}},a,b,c,d]$ are gamma
and beta so that analytic expressions are available for the
expectation and variance of the gamma sufficient statistics, as
detailed in equations \ref{eq:FirstMoment}
to \ref{eq:SecondMoment}. The key functions of $N_{s+}$ are ($a_{s}%
\prime,b_{s}\prime,c_{s}\prime,d_{s}\prime)=(a+N_{s+},b+D_{s+},c+N_{s+}%
,d+Y_{s+}-N_{s+})$ such that~:%
\[
\mathbb{E}\left(
\begin{array}
[c]{c}%
\ln(\mu_{s})\\
-\mu_{s}\\
\ln(p_{s})\\
\ln(1-p_{s})
\end{array}
|\underline{\mathbf{N}}\right)  =\left(
\begin{array}
[c]{c}%
\psi(a_{s}^{\prime})-\ln(b_{s}^{\prime})\\
-\frac{a_{s}^{\prime}}{b_{s}^{\prime}}\\
\psi(c_{s}^{\prime})-\psi(c_{s}^{\prime}+d_{s}^{\prime})\\
\psi(d_{s}^{\prime})-\psi(c_{s}^{\prime}+d_{s}^{\prime})
\end{array}
\right)
\]
and then\ the matrix $B_{s}$ is obtained by taking the covariance of
this vector.Given $\underline{N_{s}}$ additional advantage is taken
from the conditional independence of $p_{s}$ and $\mu_{s}$ (as shown
on Figure~\ref{Fig:muro} for the continuous case).
\[
\Var\left(
\begin{array}
[c]{c}%
\ln(\mu_{s})\\
-\mu_{s}\\
\ln(\rho_{s})\\
-\rho_{s}%
\end{array}
|\underline{\mathbf{N}}\right)  =\left(
\begin{array}
[c]{cccc}%
-\psi^{\prime}(a^{\prime}) & \frac{1}{b^{\prime}} & 0 & 0\\
\frac{1}{b^{\prime}} & -\frac{a^{\prime}}{b^{\prime2}} & 0 & 0\\
0 & 0 & \psi^{\prime}(c_{s}^{\prime})-\psi^{\prime}(c_{s}^{\prime}%
+d_{s}^{\prime}) & -\psi^{\prime}(c_{s}^{\prime}+d_{s}^{\prime})\\
0 & 0 & -\psi^{\prime}(c_{s}^{\prime}+d_{s}^{\prime}) & \psi^{\prime}%
(d_{s}^{\prime})-\psi^{\prime}(c_{s}^{\prime}+d_{s}^{\prime})
\end{array}
\right)
\]
and the expectation to obtain $A_{s}$ is performed via importance
sampling.

To sum it up%
\begin{equation}
I_{e}(\theta)=-\frac{\partial^{2}\ln{[\underline{\mathbf{Y}}|\theta]}%
}{\partial\theta_{i}\,\partial\theta_{j}}%
\end{equation}
At the maximum likelihood estimator $\hat{\theta}$, the following equality
occurs~:
\begin{equation}
I_{e}(\hat{\theta},\underline{\mathbf{Y}})=S\,\left(
\begin{array}
[c]{cccc}%
-\psi^{\prime}(\hat{a}) & \frac{1}{\hat{b}} & 0 & 0\\
\frac{1}{b} & -\frac{\hat{a}}{\hat{b}^{2}} & 0 & 0\\
0 & 0 & -\psi^{\prime}(\hat{c})+\psi^{\prime}(\hat{c}+\hat{d}) & \psi^{\prime
}(\hat{c}+\hat{d})\\
0 & 0 & \psi^{\prime}(\hat{c}+\hat{d}) & -\psi^{\prime}(\hat{d})+\psi^{\prime
}(\hat{c}+\hat{d})
\end{array}
\right)  +\sum_{s=1}^{S}(A_{s}+B_{s})
\end{equation}
with
\[
A_{s}=\left(
\begin{matrix}
\mathbb{E}_{\nu_{s}}(\psi^{\prime}%
(a_{s}^{\prime})) & \frac{-1}{b_{s}^{\prime}} & 0 & 0\\
\frac{-1}{b_{s}^{\prime}} & \frac{\mathbb{E}_{N_{s+}|\underline{\mathbf{Y}%
},\hat{\theta}}(a_{s}^{\prime})}{b_{s}^{\prime}{}^{2}} & 0 & 0\\
0 & 0 & \mathbb{E}_{\nu_{s}}(\psi^{\prime
}(c_{s}^{\prime})-\psi^{\prime}(c_{s}^{\prime}+d_{s}^{\prime})) &
-\mathbb{E}_{\nu_{s}}(\psi^{\prime}%
(c_{s}^{\prime}+d_{s}^{\prime}))\\
0 & 0 & -\mathbb{E}_{\nu_{s}}(\psi^{\prime
}(c_{s}^{\prime}+d_{s}^{\prime})) & \mathbb{E}_{\nu_{s}}(\psi^{\prime}(d_{s}^{\prime})-\psi^{\prime}(c_{s}^{\prime
}+d_{s}^{\prime}))
\end{matrix}
\right)
\]

and
\[
B_{s}=\Var_{N_{s+}|\hat{\theta},x}\left(
\begin{array}
[c]{c}%
\psi(a_{s}^{\prime})-\ln(b_{s}^{\prime})\\
-\frac{a_{s}^{\prime}}{b_{s}^{\prime}}\\
\psi(c_{s}^{\prime})-\psi(c_{s}^{\prime}+d_{s}^{\prime})\\
\psi(d_{s}^{\prime})-\psi(c_{s}^{\prime}+d_{s}^{\prime})
\end{array}
\right)
\]

where $a_{s}^{\prime}=\hat{a}+N_{s+}$, $b_{s}^{\prime}=\hat{b}+D_{s+}$,
$c^{\prime}=\hat{c}+N_{s+}$ and $d_{s}^{\prime}=\hat{d}+Y_{s+}-N_{s+}$ \ (
$b_{s}^{\prime}$ is the only term that is not a function of $N_{s+}$, thus
behaving like a constant with regards to the $\Var_{N_{s+}|\hat
{\theta},x}$ operator)
\section{The discrete algorithm}
\label{app:DiscAlgo}

If we adapt bluntly from the continuous version, the algoritm would
write
\begin{enumerate}
\item  Generate $N_{i}^{(g)}=0$ wherever $y_{i=0}$ for$i=I-I^{+}+1,\ldots,I$.
\item Generate a value of $N_+$ according to
$$N_+\varpropto \frac{\Gamma(a'+N_+) \Gamma(c'+N_+) \Gamma(d'+Y_+-N_+) D_+^{N_+}}{(b'+D_+)^{a'+N_+} \prod_{j=1}^{I^+} \Gamma\left(N_+\frac{Y_j D_j}{(YD)_+}\right)}$$

\item Generate each $N_i$ for $i=1,\ldots,I^+$, so that the vector $\underline{N}$ is distributed according to a multivariate hypegeometric Fisher distribution
\citep{McCullagh1983} given by

$$[\underline{N}\vert N_+]=\frac{g(\underline{N};N_+,\underline{Y},\underline{D}/D_+)}{K_{N_+}} $$
with
 $$ g(\underline{N};N_+,\underline{Y},\underline{D})=\prod_{i=I^*+1}^I \Comb{Y_j}{N_j}(D_j/D_+)^{N_j},$$
 $$K_{N_+}=\sum_{y\in \mathcal{S}} g(\underline{y};N_+,\underline{Y},\underline{D}),$$
and
 $$\mathcal{S}=\left\lbrace \underline{N}\in \Z^{I^+}_{+}\vert \sum_{i=I^*+1}^I N_i=N_+\right\rbrace.$$

 \item Associate to the vector the weight
$$w^{(g)}=K_{N^{(g)}_+}\prod_{i=I^*}^{I} \frac{\Gamma\left(N^{(g)}_+\frac{Y_j D_j}{(YD)_+}\right)}{\Gamma(N_j)}.$$
\end{enumerate}
Importance Sampling relying this time on the multivariate
hypergeometric distribution seems to stand naturally as the core of
the algorithm to evaluate (\ref{eq:condNdiscretecase}). But during
our first trials, the above adaptation of the continuous version
performed very badly, leading to a large variance of the importance
weights, i.e. a degeneracy phenomenon that would put the all weight
onto a very few contributing particles. In order to put more weight
onto particles that have a good chance to efficiently attain the
target distribution, a mixture was chosen as the importance
distribution for a modified algorithm. The idea is similar in spirit
to the auxiliary particle filtering of \cite{PittShephard99}. More
precisely, the first step consists of determining an approximate
mean of $N_{s+}$ in stratum $s$, denoted $N_{s+}^{(ref)}$. One draws
a L-sample of $N_{s+}$ according to
\[
g(N_{+})\varpropto\frac{\Gamma(a^{\prime}+N_{+})\Gamma(c^{\prime}+N_{+})\Gamma(d^{\prime}+y_{+}
-N_{+})D_{+}^{N_{+}}}{\Gamma(b^{\prime}+D_{+})^{a^{\prime}+N_{+}}\Gamma(N_{+}+1)\Gamma
(N_{+})\Gamma(Y_{+}-N_{+}+1)}%
\]
The $g$ distribution corresponds to the conditional distribution of
$\ N_{+}$ given the sum of the data collected in stratum $s$ but
ignoring the individual records$.$ $N_{s+}^{(ref)}$ is given by the
mean over a sample that is
\[
N_{s+}^{(ref)}=\frac{1}{L}\sum N_{+}^{(i)}%
\]
and provides a good estimation of the location\ of $N_{s+}$. As
previously we omit the index $s$ to make the reading easier.
Subsequently, the following algorithm relies on independent but non
identically distributed simulations~:

\begin{enumerate}
\item Generate $N_i^{(g)}=0$ wherever $y_i=0$ for $i=I-I^++1,\ldots,I$.
\item  Draw $\mu^{(g)}\sim\Gamma(a^{\prime}+N_{+}^{(ref)},b^{\prime}+D_{+})$ and $p^{(g)}\sim\beta(c^{\prime}+ N_{s+}^{(ref)}, d^{\prime}+Y_{s+} -N_{s+}^{(ref)})$

\item  Given $\mu^{(g)}$ and $\rho^{(g)}$, draw $N_{sk}^{(g)}\sim[N_{sk}%
\vert\mu^{(g)}, p^{(g)}, y_{sk}]$ that is~:
$$[N_{i}^{(g)}=k]=K_i\left(\frac{\mu^{(g)} p^{(g)} D_i}{1-p^{(g)}}\right)^{N_i}
\frac{1}{\Gamma(N_i)\Gamma(Y_i-N_i+1)\Gamma(N_i+1)}  \1{0<N_{sk}\leq
Y_{si}} , $$ where $K_i$ denotes the normalizing constant.

\item  Compute the weight of each particle $g$ using
\[
w^{(g)}=\prod_{i=1}^{I^+} \frac{\Gamma(N_i+1)}{K_i (\mu^{(g)}
p^{(g)})^{N_i}}\left(
\frac{\Gamma(a^{\prime}+N_{s+})\Gamma(N_{s+}+c^{\prime})\Gamma
    (Y_{+}-N_{s+}+d^{\prime})}{(b^{\prime}+D_{s+})^{N_{s+}}}\right)
\]
\end{enumerate}

\end{document}